\documentclass[11pt,twoside,leqno]{aomamlt2e} 
 
\pageno{1}
\received{August 2, 2005}

\theoremstyle{plain}
\newtheorem{theorem}{Theorem}[section]
\newtheorem{proposition}[theorem]{Proposition}
\newtheorem{lemma}[theorem]{Lemma}
\newtheorem{corollary}[theorem]{Corollary}
\newtheorem{assumption}[theorem]{Assumption}
\newtheorem{definition}[theorem]{Definition}
\newtheorem{remark}[theorem]{Remark}
\newtheorem{sublemma}[theorem]{Sublemma}

\newenvironment{remarks}{\begin{myremarks}\begin{nummer}}%
    {\end{nummer}\end{myremarks}}

\newtheorem{myremarks}[theorem]{Remark}

\newcounter{numcount}
\newcommand{\labelnummer}{\mbox{\normalfont (\roman{numcount})}}%

\makeatletter

\newenvironment{nummer}%
  {\let\curlabelspeicher\@currentlabel%
    \begin{list}{\labelnummer}%
      {\usecounter{numcount}\leftmargin0pt%
        \topsep0.5ex\partopsep2ex\parsep0pt\itemsep0ex\@plus1\p@%
        \labelwidth2.5em\itemindent3.5em\labelsep1em%
      }%
    \let\saveitem\item%
    \def\item{\saveitem%
      \def\@currentlabel{\curlabelspeicher$\,$\labelnummer}}%
    \let\savelabel\label%
    \def\label##1{\savelabel{##1}%
      \@bsphack%
        \ifmmode\else%
          \protected@write\@auxout{}%
          {\string\newlabel{##1item}{{\labelnummer}{\thepage}}}%
        \fi%
      \@esphack%
    }%
  }{\end{list}}%

\def\itemref#1{\expandafter\@setref\csname r@#1item\endcsname%
  \@firstoftwo{#1}}%

\newif\ifper\pertrue
\def\per{}

\def\au#1#2{#2, #1}
\def\lau#1#2{#2, #1:}
\def\et{,\ }
\def\ti#1{{#1}.}

\def\bti{\@ifnextchar[\bbti\bbbti}
\def\bbti[#1]#2{{#2}. #1.}
\def\bbbti#1{{#1}.}

\def\z{\@ifnextchar[\zz\zzz}
\def\zz[#1]#2#3#4#5{\perfalse{#2} \textbf{#3}, #4 (#5) [#1]\per}
\def\zzz#1#2#3#4{{#1} \textbf{#2}, #3 (#4)\ifper\per\fi\pertrue}

\def\pub{\@ifstar\pubstar\pubnostar}
\def\pubnostar{\@ifnextchar[\@@pubnostar\@pubnostar}
\def\@@pubnostar[#1]#2#3#4{#3: #2 #4, #1\per}
\def\@pubnostar#1#2#3{#2: #1 #3\ifper\per\fi\pertrue}
\def\pubstar[#1]#2#3#4{\perfalse #3: #2 #4 [#1]\per}

\renewcommand{\theequation}{\thesection.\arabic{equation}}

\let\isave\i
\def\i{\ifmmode%
           \@ifundefined{comp}%
                 {\mathrm{i}}%
                 {\hspace{0.07em}\mathrm{i}\hspace{0.07em}}%
       \else\isave%
       \fi
}
\def\e{\mathrm{e}}
\def\d{\mathrm{d}}
\def\Chi{\raisebox{.4ex}{$\chi$}}
\def\Re{\mathop\mathrm{Re}}
\def\Im{\mathop\mathrm{Im}}

\def\dist{\mathop\mathrm{dist}}

\def\tr{\mathop\mathrm{tr}}

\DeclareMathOperator{\supp}{supp}

\providecommand{\varkappa}{\kappa}

\def\llangle{\langle\mkern-4mu\langle}
\def\rrangle{\rangle\mkern-4mu\rangle}

\def\le{\leqslant}
\def\ge{\geqslant}
\def\emptyset{\varnothing}

\newcommand{\EE}{\mathbb{E}}

\newcommand{\PP}{\mathbb{P}}
\newcommand{\RR}{\mathbb{R}}
\newcommand{\Sbb}{\mathbb{S}}
\newcommand{\ZZ}{\mathbb{Z}}
\newcommand{\E}{\mathbb{E}}

\def\A{\mathbf{A}}

\def\cE{\mathcal{E}}
\def\cH{\mathcal{H}}
\def\cK{\mathcal{K}}
\def\cL{\mathcal{L}}
\def\cT{\mathcal{T}}
\def\cU{\mathcal{U}}
\def\cP{\mathcal{P}}
\def\cJ{\mathcal{J}}

\def\cQ{\mathcal{Q}}

\newcommand{\tnorm}[1]{\left|\!\left|\!\left| #1 \right|\!\right|\!\right|}
\newcommand{\norm}[1]{\lVert #1 \rVert}
\newcommand{\abs}[1]{\lvert #1 \rvert}
 
\newcommand{\hnorm}[1]{\left\{ \!\left\{ #1\right\}\! \right\}}

\renewcommand{\currannalsline}[2]{\titlepage
  \vglue-42pt\centerline{\hfill{\scriptsize To appear in the Annals of
      Mathematics} \hfill}} 

\begin{document}
\currannalsline{XXX}{XXX} 

\title{On Mott's formula for the ac-conductivity\\ in the Anderson model}

\acknowledgement{A.K.\ was supported in part by NSF Grant DMS-0457474.
  P.M.\ was supported by the Deutsche Forschungsgemeinschaft
  (DFG) under grant Mu~1056/2--1.}

\twoauthors{Abel Klein, Olivier Lenoble,}{Peter M\"uller}

\institution{University of California, Irvine, California, USA\\
  \email{aklein@uci.edu}\\ 
  \email{lenoble@math.uci.edu}\\
  \email{peter.mueller@physik.uni-goe.de} \\[1ex] 
  {(\normalfont P.M.\ was on leave from the} Institut f\"ur
  Theoretische Physik, Georg-August-Universit\"at, G\"ottingen, Germany)}

\shorttitle{On Mott's formula for the ac-conductivity}

\begin{abstract} 
  We study the ac-conductivity in linear response theory in the general
  framework of ergodic magnetic Schr\"odinger operators.  For the Anderson
  model, if the Fermi energy lies in the localization regime, we prove that
  the ac-conductivity is bounded by $ C \nu^2 (\log \frac 1 \nu)^{d+2}$ at
  small frequencies $\nu$. This is to be compared to Mott's formula, which
  predicts the leading term to be $ C \nu^2 (\log \frac 1 \nu)^{d+1}$.
\end{abstract}

%
\section{Introduction}
The occurrence of localized electronic states in disordered systems was first
noted by Anderson in 1958 \cite{And58}, who argued that for a simple
Schr\"odinger operator in a disordered medium,``at sufficiently low densities
transport does not take place; the exact wave functions are localized in a
small region of space."  This phenomenon was then studied by Mott, who wrote
in 1968 \cite{Mot68}: ``The idea that one can have a continuous range of
energy values, in which all the wave functions are localized, is surprising
and does not seem to have gained universal acceptance.''  This led Mott to
examine Anderson's result in terms of the Kubo--Greenwood formula for
$\sigma_{E_{F}}(\nu)$, the electrical alternating current (ac) conductivity at
Fermi energy $E_{F}$ and zero temperature, with $\nu$ being the frequency.
Mott used its value at $\nu=0$ to reformulate localization: If a range of values of
the Fermi energy $E_{F}$ exists in which $ \sigma_{E_{F}}(0) =0$, the states
with these energies are said to be localized; if $ \sigma_{E_{F}}(0) \not=0$,
the states are non-localized.

Mott then argued that the direct current (dc) conductivity
$\sigma_{E_{F}}(0)$ indeed vanishes in the localized regime. In the context of
Anderson's model, he studied the behavior of $\Re \sigma_{E_{F}}(\nu) $ as
$\nu \to 0$ at Fermi energies $E_{F}$ in the localization region (note $\Im
\sigma_{E_{F}}(0)=0$).  The result was the well-known \emph{Mott's formula}
for the ac-conductivity at zero temperature \cite{Mot68,Mot70}, which we state
as in \cite[Eq.~(2.25)]{MotD} and \cite[Eq.~(4.25)]{LiGr88}:
\begin{equation}
  \label{mott}
 \Re \sigma_{E_{F}}(\nu)  \sim  n({E_{F}})^{2} \, \tilde{\ell}_{E_{F}}^{d+2}\,
  \nu^{2} 
  \left(\log\tfrac{1}{\nu}\right)^{d+1}  \quad \text{as} \quad  \nu \downarrow
  0, 
\end{equation}
where $d$ is the space dimension, $n({E_{F}})$ is the density of states at
energy $E_{F}$, and $\tilde{\ell}_{E_{F}}$ is a localization length at energy
$E_{F}$.
  
Mott's calculation was based on a fundamental assumption: the leading
mechanism for the ac-conductivity in localized systems is the resonant
tunneling between pairs of localized states near the Fermi energy $E_{F}$,  the transition from  a state of energy $E \in ]E_{F}-{\nu},E_{F}] $ to another state with  resonant energy $E+\nu$, the energy for the
transition being provided by the electrical field.   Mott also argued that the two resonating states must be  located at a spatial distance of $\sim \log \frac{1}{\nu}$. Kirsch, Lenoble and Pastur
\cite{KiLe03} have recently provided a careful heuristic derivation of Mott's formula
along these lines, incorporating also ideas of Lifshitz \cite{Lif65}.

In this article we give the first mathematically rigorous treatment of Mott's
formula.  The general nature of Mott's arguments leads to the belief in 
physics  that Mott's formula \eqref{mott} describes the generic
behavior of the low-frequency conductivity in the localized regime,
irrespective of model details.  Thus we study it in the most popular model for
electronic properties in disordered systems, the Anderson tight-binding model
\cite{And58} (see \eqref{AND}), where we prove a result of the form
\begin{equation}
  \label{leadingbound}
  \Re  {\overline{\sigma}_{E_{F}}(\nu)  } 
   \le { \,c \;}
  \tilde{\ell}_{E_{F}}^{d+2}\,
  \nu^{2} \left(\log \tfrac{1}{\nu}\right)^{d+2}  \quad \text{for small
    $\nu>0$}. 
\end{equation}
The precise result is stated in Theorem~\ref{ourmott}; formally  
\begin{equation} 
  \Re  {\overline{\sigma}_{E_{F}}(\nu)  }=\frac 1 \nu \int_{0}^{\nu} \d
  \nu^{\prime}\,  \Re  \sigma_{E_{F}}(\nu^{\prime}) ,
\end{equation} 
so $\Re  {\overline{\sigma}_{E_{F}}(\nu)  } \approx   \Re
  {{\sigma}_{E_{F}}(\nu)}$ for small $\nu >0$.
The discrepancy in the exponents of $\log\frac{1}{\nu}$ in
\eqref{leadingbound} and \eqref{mott}, namely $d+2$ instead of $d+1$, is
discussed in Remarks~\ref{remMot} and \ref{missingpower2}, where we give
arguments in support of $d+2$.

We believe that a result similar to Theorem~\ref{ourmott} holds for the
continuous Anderson Hamiltonian, which is a random Schr\"odinger operator on
the continuum with an alloy-type potential.  All steps in our proof of
Theorem~\ref{ourmott} can be redone for such a continuum model, except the
finite volume estimate of Lemma~\ref{minami}.  The missing ingredient is
Minami's estimate \cite{Min96}, which we recall in \eqref{minami2}.  It is not yet
available for that continuum model.  In fact, proving a continuum analogue of
Minami's estimate would not only yield Theorem~\ref{ourmott} for the
continuous Anderson Hamiltonian, but it would also establish, in the
localization region, simplicity of eigenvalues as in \cite{KlMo05} and Poisson
statistics for eigenvalue spacing as in \cite{Min96}.
  
To get to Mott's formula,  we conduct what seems to be the first careful mathematical analysis of the ac-conductivity in linear response theory,  and  introduce a new concept,  the conductivity measure.  This is done
in the general framework of ergodic magnetic Schr\"odinger operators,
in both  the discrete and continuum settings.  We give a controlled derivation
in linear response theory of a Kubo formula for the ac-conductivity along the
lines of the derivation for the dc-conductivity given in \cite{BGKS}. This
Kubo formula (see Corollary~\ref{corSigma}) is written in terms of
$\Sigma_{E_{F}}(\d \nu)$, the conductivity measure at Fermi energy $E_{F}$
(see Definition~\ref{defSigma} and Theorem~\ref{thmSigma}).  If $\Sigma_{E_{F}}(\d \nu)$ was known to be an absolutely continuous measure, $ \Re \sigma_{E_{F}}(\nu)$ would then be well-defined as its density.  The conductivity measure $\Sigma_{E_{F}}(\d \nu)$ is thus an analogous concept to the density of states measure $\mathcal{N}(\d E)$, whose formal density is the density of states $n(E)$.
The conductivity measure has also an expression in terms of the velocity-velocity correlation measure (see Proposition~\ref{PropSigmaPhi}).

The first mathematical proof of localization \cite{GMP} appeared almost twenty
years after Anderson's seminal paper \cite{And58}.  This first mathematical
treatment of Mott's formula is appearing about thirty seven years after its
formulation \cite{Mot68}. It relies on some highly nontrivial research on
random Schr\" odinger operators conducted during the last thirty years, using
a good amount of what is known about the Anderson model and localization.  The
first ingredient is linear response theory for ergodic Schr\" odinger
operators with Fermi energies in the localized region \cite{BGKS}, from which
we obtain an expression for the conductivity measure.  To estimate the low
frequency ac-conductivity, we restrict the relevant quantities to finite
volume and estimate the error.  The key ingredients here are the
Helffer--Sj\"ostrand formula for smooth functions of self-adjoint operators
\cite{HeSj89} and the exponential estimates given by the fractional moment
method in the localized region \cite{AiMo93,Aiz94,AiSc01}.  The error
committed in the passage from spectral projections to smooth functions is
controlled by Wegner's estimate for the density of states \cite{Weg81}.  The
finite volume expression is then controlled by Minami's estimate \cite{Min96},
a crucial ingredient.  Combining all these estimates, and choosing the size of
the finite volume to optimize the final estimate, we get \eqref{leadingbound}.

This paper is organized as follows.  In Section~\ref{sec:main} we introduce
the Anderson model, define the region of complete localization, give a brief
outline of how electrical conductivities are defined and calculated in linear
response theory, and state our main result (Theorem~\ref{ourmott}).  In
Section~\ref{sec:Kubo}, we give a detailed account of how electrical
conductivities are defined and calculated in linear response theory, within
the noninteracting particle approximation.  This is done in the general
framework of ergodic magnetic Schr\"odinger operators; we treat simultaneously
the discrete and continuum settings.  We introduce and study the conductivity
measure (Definition~\ref{defSigma}), and derive a Kubo formula
(Corollary~\ref{corSigma}).  In Section~\ref{sec:mottproof} we give the proof
of Theorem~\ref{ourmott}, reformulated as Theorem~\ref{ourmott2}.

In this article $\abs{B} $ denotes either Lebesgue measure if $B$ is a
Borel subset of $\RR^{n}$, or the counting measure if $B\subset\ZZ^{n}$
($n=1,2,\ldots$). We always use $\Chi_{B}$ to denote the characteristic function of the set $B$.  By $C_{a,b, \ldots}$, etc., will always denote some finite
constant depending only on $a,b, \ldots$.

%
\section{The Anderson model and the main result}
\label{sec:main}
The Anderson tight binding model is described by the random Schr\"odinger
operator $H$, a measurable map $\omega \mapsto H_{\omega}$ from a probability
space $(\Omega,\mathbb{P})$ (with expectation $\mathbb{E}$) to bounded
self-adjoint operators on $\ell^2(\ZZ^d)$, given by
\begin{equation}
  \label{AND}
  H_\omega := - \Delta + V_\omega .
\end{equation} 
Here $\Delta$ is the centered discrete  Laplacian, 
\begin{equation}
 (\Delta \varphi)(x):= -  \sum_{{y\in\ZZ^d; \, |x-y|=1}} \varphi(y) \quad
 \text{for} \quad   \varphi\in\ell^2(\ZZ^d), 
\end{equation}
and the random potential $V$ consists of independent identically distributed
random variables $\{V(x) ; x \in \ZZ^d\}$ on $(\Omega,\mathbb{P})$, such that
the common single site probability distribution $\mu$ has a bounded density
$\rho$ with compact support.

The \emph{Anderson Hamiltonian} $H$ given by \eqref{AND} is
$\mathbb{Z}^d$-ergodic, and hence its spectrum, as well as its spectral
components in the Lebesgue decomposition, are given by non-random sets
$\PP$-almost surely \cite{KiMa82, CaLa90, PaFi92}.

There is a wealth of localization results for the Anderson model in arbitrary
dimension, based either on the multiscale analysis \cite{FrSp83, FrMa85,Sp,
  DrKl89}, or on the fractional moment method \cite{AiMo93, Aiz94, AiSc01}.
The spectral region of applicability of both methods turns out to be the same,
and in fact it can be characterized by many equivalent conditions
\cite{GKduke,GKjsp}.  For this reason we call it the \emph{region of complete
  localization} as in \cite{GKjsp}; the most convenient definition for our
purposes is by the conclusions of the fractional moment method.

\begin{definition}
  \label{locdef}
  The region of complete localization $\Xi^{\mathrm{CL}}$ for the Anderson
  Hamiltonian $H$ is the set of energies $E \in \RR$ for which there is an
  open interval $I_{E} \ni E$ and an exponent $s=s_{E}\in]0,1[$  such that
  \begin{equation}
    \label{aizenmolch}
    \sup_{E^{\prime} \in I_{E}}  \sup_{ \eta \not=0} \EE\bigl\{
    |\langle\delta_x,R(E^{\prime} +\i\eta) \delta_y \rangle |^{s} \bigr\}\le
    K\, \e^{- \frac 1 {\ell} \abs{x-y}} 
    \quad \text{for all} \;  x,y\in\ZZ^d,
  \end{equation}  
  where $K=K_{E}$ and $\ell=\ell_{E} >0$ are constants, and $R(z):=(H-z)^{-1}$
  is the resolvent of $H$.
\end{definition}

\begin{remarks}
\item 
   The constant $\ell_{E}$ admits the interpretation of a localization
  length at energies near  $E$.
\item 
  The fractional moment condition \eqref{aizenmolch} is known to
  hold under various circumstances, for example,  large
  disorder or extreme energies \cite{AiMo93,  Aiz94, 
  AiSc01}. Condition \eqref{aizenmolch} implies spectral localization
  with exponentially decaying eigenfunctions \cite{AiMo93}, dynamical
  localization \cite{Aiz94, AiSc01}, exponential decay of the Fermi
  projection \cite{AiGr98}, and  absence of level repulsion \cite{Min96}.   
\item   
  The single site potential density $\rho$ is assumed to be bounded with
  compact support, so condition \eqref{aizenmolch} holds with any exponent $ s
  \in ]0,\frac 1 4[ $ and appropriate constants $K(s)$ and $\ell(s)>0$ at all
  energies where a multiscale analysis can be performed \cite{AiSc01}.  Since
  the converse is also true, that is, given \eqref{aizenmolch} one can perform
  a multiscale analysis as in \cite{DrKl89} at the energy $E$, the energy
  region $\Xi^{\mathrm{CL}}$ given in Definition~\ref{locdef} is the same
  region of complete localization defined in \cite{GKjsp}.
\end{remarks}

We briefly outline how electrical conductivities are defined and
calculated in linear response theory following the approach adopted in
\cite{BGKS}; a detailed account in the general framework of ergodic
magnetic Schr\"odinger operators, in both the discrete and continuum
settings, is given in Section~\ref{sec:Kubo}.

Consider a system at zero temperature, modeled by the
Anderson Hamiltonian $H$.  At the reference time $t=-\infty$, the system is in
equilibrium in the state given by the (random) Fermi projection
$P_{E_{F}}:=\Chi_{]-\infty,E_{F}]}(H)$, where we assume that $E_{F}\in
\Xi^{\mathrm{CL}}$, that is, the Fermi energy lies in the region of complete
localization.  A spatially homogeneous, time-dependent electric field
$\mathbf{E}(t)$ is then introduced adiabatically: Starting at time $t=
-\infty$,  we switch on the
electric field $\mathbf{E}_{\eta}(t):= \e^{\eta t}\mathbf{E}(t)$ with $\eta
>0$, and then let $\eta \to 0$. On account of isotropy we assume without
restriction that the electric field is pointing in the $x_{1}$-direction:
$\mathbf{E}(t)=\mathcal{E}(t) \widehat{x}_{1} $, where $\mathcal{E}(t)$ is the
(real-valued) amplitude of the electric field, and $\widehat{x}_{1}$ is the
unit vector in the $x_{1}$-direction. We assume that
\begin{equation}\label{FT}
  \cE(t) =  \int_{\RR}\!\d \nu \; \e^{\i\nu t}\widehat{\cE}(\nu),  \quad
  \text{where}   \  \widehat{\cE}(\nu) \in C_{c}(\RR) \ \text{and} \
  \widehat{\cE}(\nu)=\overline{ \widehat{\cE}(-\nu)}. 
\end{equation}
For each $\eta>0$ this results in a time-dependent random Hamiltonian
$H({\eta, t}) $, written in an appropriately chosen gauge.  The system is then
described at time $t$ by the density matrix $\varrho(\eta,t)$, given as the
solution to the Liouville equation
\begin{equation}\label{Liouvilleeq}
  \left\{
    \begin{array}{l}i\partial_t \varrho(\eta,t) = [H(\eta,t),\varrho(\eta,t)]
      \\
      \lim_{t \to  -\infty}\varrho(\eta,t)= P_{E_{F}} \end{array}
  \right.  .
\end{equation}
The adiabatic electric field generates a time-dependent electric current,
which, thanks to reflection invariance in the other directions, is also
oriented along the $x_{1}$-axis, and has amplitude
\begin{equation}
  \label{curdef}
  J_{\eta}(t;E_{F},\mathcal{E}) = -  \cT \bigl( \varrho(\eta,t)
  \dot{X}_{1}(t)\bigr)\,,  
\end{equation}
where $\cT $ stands for the trace per unit volume and $\dot{X}_{1}(t)$ is the
first component of the velocity operator at time $t$ in the Schr\"odinger
picture (the time dependence coming from the particular gauge of the
Hamiltonian).  In Section~\ref{sec:Kubo} we calculate the \emph{linear
  response current}
\begin{equation}
  \label{lincur2}
  J_{\eta,\mathrm{lin}}(t;E_{F},\cE) := \frac{\d}{\d\alpha}\,
  J_{\eta }(t; E_{F},\alpha\cE)\big|_{\alpha=0}.
\end{equation}
The resulting Kubo formula  may be written    as
\begin{equation}
  \label{lincur-rig}
  J_{\eta,\mathrm{lin}}(t; E_F, \cE) = \e^{\eta t} \int_\RR\!\d\nu\;
  \e^{\i\nu t}\, \sigma_{E_{F}}(\eta,\nu) \, \widehat{\cE}(\nu),
\end{equation}
with the (regularized) \emph{conductivity} $ \sigma_{E_{F}}(\eta,\nu)$ given
by
\begin{equation}
  \label{regcond}
  \sigma_{E_{F}}(\eta,\nu):=- \tfrac{\i}\pi \int_{\RR} \! \Sigma_{E_{F}}(\d
  \lambda)  \left(\lambda  + \nu - \i \eta  \right)^{-1} , 
\end{equation}
where $\Sigma_{E_{F}}$ is a finite, positive, even Borel measure on $\RR$, the
\emph{conductivity measure} at Fermi Energy ${E_{F}}$---see
Definition~\ref{defSigma} and Theorem~\ref{thmSigma}.
    
It is customary to decompose $ \sigma_{E_{F}}(\eta,\nu)$ into its real and
imaginary parts:
\begin{equation}
  \sigma_{E_{F}}^{\mathrm{in}}(\eta,\nu):= \Re  \sigma_{E_{F}}(\eta,\nu) \quad 
  \text{and} \quad  \sigma_{E_{F}}^{\mathrm{out}}(\eta,\nu):= \Im
  \sigma_{E_{F}}(\eta,\nu), 
\end{equation}
the \emph{in phase} or \emph{active} conductivity $
\sigma_{E_{F}}^{\mathrm{in}}(\eta,\nu)$ being an even function of $\nu$, and
the \emph{out of phase} or \emph{passive} conductivity
$\sigma_{E_{F}}^{\mathrm{out}}(\eta,\nu)$ an odd function of $\nu$.  This
induces a decomposition $ J_{\eta,\mathrm{lin}} =
J_{\eta,\mathrm{lin}}^{\mathrm{in}} + J_{\eta,\mathrm{lin}}^{\mathrm{out}}$ of
the linear response current into an \emph{in phase} or \emph{active}
contribution
\begin{align}
  \label{lincur-rig-in}
  J_{\eta,\mathrm{lin}}^{\mathrm{in}}(t; E_F, \cE) & := \e^{\eta t}
  \int_\RR\!\d\nu\; \e^{\i\nu t}\, \sigma_{E_{F}}^{\mathrm{in}}(\eta,\nu) \,
  \widehat{\cE}(\nu), 
  \intertext{and an  \emph{out of phase} or \emph{passive} contribution}
  J_{\eta,\mathrm{lin}}^{\mathrm{out}}(t; E_F, \cE) & :=\i \, \e^{\eta t}
  \int_\RR\!\d\nu\; 
  \e^{\i\nu t}\, \sigma_{E_{F}}^{\mathrm{out}}(\eta,\nu) \,
  \widehat{\cE}(\nu). 
\end{align}
The adiabatic limit $\eta \downarrow 0$ is then performed, yielding
\begin{equation}
J_{\mathrm{lin}}(t; E_F,\cE)=J_{\mathrm{lin}}^{\mathrm{in}}(t; E_F,
\cE)+J_{\mathrm{lin}}^{\mathrm{out}}(t; E_F, \cE).
\end{equation}  
In particular we obtain the following expression for the linear response in
phase current (see Corollary~\ref{corSigma}):
\begin{align}
  \label{in-rig2}
  J_{\mathrm{lin}}^{\mathrm{in}}(t; E_F, \cE):= \lim_{\eta\downarrow 0}
  J_{\eta,\mathrm{lin}}^{\mathrm{in}}(t; E_F, \cE) =
  \int_\RR\!\Sigma_{E_{F}}(\d\nu)\; \e^{\i\nu t}\, \widehat{\cE}(\nu).
\end{align}
The terminology comes from the fact that if the time dependence of the
electric field is given by a pure sine (cosine), then $
J_{\mathrm{lin}}^{\mathrm{in}}(t; E_F, \cE)$ also varies like a sine (cosine)
as a function of time, and hence is in phase with the field, while
$J_{\mathrm{lin}}^{\mathrm{out}}(t; E_F, \cE)$ behaves like a cosine (sine),
and hence is out of phase. Thus the work done by the electric field on the
current $J_{\mathrm{lin}}(t; E_F, \cE)$ relates only to
$J_{\mathrm{lin}}^{\mathrm{in}}(t; E_F, \cE)$ when averaged over a period of
oscillation. The passive part $J_{\mathrm{lin}}^{\mathrm{out}}(t; E_F, \cE)$
does not contribute to the work.
  
It turns out that the in phase conductivity
\begin{equation}
\sigma_{E_{F}}^{\mathrm{in}}(\nu)=\Re \sigma_{E_{F}}(\nu) :=\lim_{{\eta
    \downarrow 0}} \sigma_{E_{F}}^{\mathrm{in}}(\eta,\nu),
    \end{equation}
     appearing in Mott's
formula \eqref{mott}, and more generally in physics  (e.g.,
\cite{LiGr88,KiLe03}), may not be a well defined
function.  It is the conductivity measure $\Sigma_{E_{F}}$ that is a well
defined mathematical quantity.  If the measure $\Sigma_{E_{F}}$ happens to be
absolutely continuous, then the two are related by
$\sigma_{E_{F}}^{\mathrm{in}}(\nu):= \frac {\Sigma_{E_{F}}(\d\nu)}{\d\nu}$,
and \eqref{in-rig2} can be recast in the form
\begin{equation}
  \label{linresp2}
  J_{\mathrm{lin}}^{\mathrm{in}}(t; E_F, \cE) = \int_{\RR}\!\d\nu \; \e^{\i\nu
  t} \, \sigma_{E_{F}}^{\mathrm{in}}(\nu) \; \widehat{\cE}(\nu).
\end{equation}

Since the in phase conductivity $\sigma_{E_{F}}^{\mathrm{in}}(\nu)$ may not be
well defined as a function, we state our result in terms of the \emph{average
  in phase conductivity}, an even function ($  \Sigma_{E_{F}}$ is an even measure) defined by
\begin{equation} \label{overlinesigma}
  \overline{\sigma}_{E_{F}}^{\mathrm{in}}(\nu)
  := \tfrac 1 \nu\,
  \Sigma_{E_{F}}([0,\nu]) \quad \text{for $\nu>0$}.
  \end{equation}

Our main result is given in the following theorem, proven in
Section~\ref{sec:mottproof}.

\begin{theorem}
  \label{ourmott}
  Let $H$ be the Anderson Hamiltonian and consider a Fermi energy in its
  region of complete localization: $E_{F} \in \Xi^{\mathrm{CL}}$. Then
  \begin{equation}
    \label{leadingbound8}
    \limsup_{\nu \downarrow 0} \frac {
      \overline{\sigma}_{E_{F}}^{\mathrm{in}}(\nu)} 
    { \nu^{2} \left(\log\tfrac{1}{\nu}\right)^{d+2}} \le  {C}^{d+2}\pi^{3}\,
    \|\rho\|_\infty^2 {\ell}_{E_{F}}^{d+2} ,
  \end{equation}
  where $\ell_{E_{F}}$ is given in \eqref{aizenmolch}, $\rho$ is the density
  of the single site potential, and the constant $C$ is independent of all
  parameters.
\end{theorem}

\begin{remark}\label{remMot1}
  The estimate \eqref{leadingbound8} is the first mathematically rigorous
  version of Mott's formula \eqref{mott}.  The proof in
  Section~\ref{sec:mottproof} estimates the constant: $C \le 205$; tweaking
  the proof would improve this numerical estimate to $C \le 36$.  The length $
  {\ell}_{E_{F}} $, which controls the decay of the $s$-th fractional moment
  of the Green's function in \eqref{aizenmolch}, is the effective localization
  length that enters our proof and, as such, is analogous to
  $\tilde{\ell}_{E_{F}}$ in \eqref{mott}.  The appearance of the term $
  \|\rho\|_\infty^2$ in \eqref{leadingbound8} is also compatible with
  \eqref{mott} in view of Wegner's estimate \cite{Weg81}: $n({E})\le
  \|\rho\|_\infty$ for a.e.\ energy $E \in \RR$.
\end{remark}

\begin{remark}\label{remMot}
  A comparison of the estimate \eqref{leadingbound8} with the expression in
  Mott's formula \eqref{mott} would note the difference in the power of
  $\log\frac{1}{\nu}$, namely $d+2$ instead of $d+1$.  This comes from a
  finite volume estimate (see Lemma~\ref{minami}) based on a result of Minami
  \cite{Min96}, which tells us that we only need to consider pairs of resonating localized states
 with energies $E$ and $E+\nu$ in a volume of diameter $\sim
  \log\frac{1}{\nu}$, which gives a factor of $ (\log\frac{1}{\nu})^{d}$.  On
  the other hand, Mott's argument \cite{Mot68,Mot70,MotD,KiLe03} assumes that
  these localized states must be at a distance $\sim \log\frac{1}{\nu}$ from
  each other, which only gives a surface area factor of $
  (\log\frac{1}{\nu})^{d-1}$.
  We have not seen any convincing argument for Mott's assumption. (See
  Remark~\ref{missingpower2} for a more precise analysis based on the proof of
  Theorem~\ref{ourmott}.)
\end{remark}

\begin{remark}
  A zero-frequency (or dc) conductivity at zero temperature may also be
  calculated by using a constant (in time) electric field.  This
  dc-conductivity is known to exist and to be equal to zero for the Anderson
  model in the region of complete localization
  \cite[Theorem~1.1]{Nak02},\cite[Corollary~5.12]{BGKS}.
\end{remark}

%
\section{Linear response theory and the conductivity measure}
\label{sec:Kubo}
In this section we study the ac-conductivity in linear response theory and
introduce the conductivity measure.  We work in the general framework of
ergodic magnetic Schr\"odinger operators, following the approach in
\cite{BGKS}. (See  \cite{BESB,SBB} for an approach incorporating dissipation.)
We treat simultaneously the discrete and continuum settings. But we will
concentrate on the zero temperature case for simplicity, the general case
being not very different.

%
\Subsec{Ergodic magnetic Schr\"odinger operators}  
We consider an ergodic magnetic Schr\"odinger operator $H$ on the Hilbert
space $\cH$, where $\cH= \mathrm{L}^2{(\RR^{d}})$ in the continuum setting and
$\cH=\ell^{2}(\ZZ^{d})$ in the discrete setting. In either case $\cH_{c}$
denotes the subspace of functions with compact support.  The ergodic operator $H$ is a
measurable map from the probability space $(\Omega, \PP)$ to the self-adjoint
operators on $\cH$. The probability space $(\Omega, \PP)$ is equipped with an
ergodic group $\{\tau_{a}; \ a \in \ZZ^d\}$ of measure preserving
transformations. The crucial property of the ergodic system is that it
satisfies a covariance relation: there exists a unitary projective
representation $U(a)$ of $\ZZ^d$ on $\cH$, such that for all $a ,b \in \ZZ^d$
and $\PP$-a.e.\ $\omega \in \Omega$ we have
\begin{align}
 \label{covintro}
    U(a) H_\omega U(a)^* &= H_{\tau_{a}(\omega)}  ,\\ \label{covintro2}
    U(a) \Chi_b U(a)^* &=\Chi_{b+a}, \\
    U(a) \delta_b &=\delta_{b+a}  \quad \text{if} \quad
    \cH=\ell^{2}(\ZZ^{d}) ,  \label{covintro3}
\end{align}
where $\Chi_a$ denotes the multiplication operator by the characteristic
function of a unit cube centered at $a$, also denoted by $\Chi_{a}$.  In
  the discrete setting the operator $\Chi_{a}$ is just the orthogonal
  projection onto the one-dimensional subspace spanned by $\delta_{a}$; in particular, 
  \eqref {covintro2} and   \eqref {covintro3} are equivalent in the discrete setting.

We assume the  ergodic magnetic Schr\"odinger operator  to be of the form
\begin{equation}  \label{Homega}
  H_\omega=\left\{
    \begin{array}{l@{\:}ll} H(\A_{\omega},V_{\omega})
            &:= \left(- \i\,\nabla - \A_\omega\right)^2 +  V_\omega \quad 
            &\text{if} \quad \cH=\mathrm{L}^{2}(\RR^{d}) \\
         H({ {\vartheta}_{\omega}},V_{\omega})
            &:= -\Delta({{\vartheta}_{\omega}}) + V_\omega 
            & \text{if} \quad \cH=\ell^{2}(\ZZ^{d})
    \end{array}\right.   . 
\end{equation}
The precise requirements in the continuum are described in \cite[
Section~4]{BGKS}.  Briefly, the random magnetic potential $\textbf{A}$ and the
random electric potential $V$ belong to a very wide class of
potentials which ensures that $H(\A_{\omega},V_{\omega})$ is essentially
self-adjoint on $\mathcal{C}_c^\infty(\RR^d)$ and uniformly bounded from below
for $\PP$-a.e.\ $\omega$, and hence there is $\gamma \ge 0$ such that
\begin{equation}\label{lowerbound}
  H_\omega +\gamma \ge 1 \quad \text{for $\PP$-a.e.\  $\omega$}.
\end{equation}
In the discrete setting ${ {\vartheta}}$ is a lattice random magnetic
potential and we require the random electric potential $V$ to be
$\PP$-almost surely bounded from below.  Thus, if we let
$\mathcal{B}(\ZZ^{d}) :=\{ (x,y) \in \ZZ^{d}\times\ZZ^{d}; |x-y|=1\}$, the set
of oriented bonds in $\ZZ^{d}$, we have ${ {\vartheta}_{\omega}}\colon
\mathcal{B}(\ZZ^{d})\to \RR$, with $ { {\vartheta}_{\omega}}(x,y)=- {
  {\vartheta}_{\omega}}(y,x)$ a measurable function of $\omega$, and
\begin{equation}
  \bigl(\Delta({ {\vartheta}_{\omega}})\varphi \bigr)(x):= -
  \sum_{{y\in\ZZ^d; \, |x-y|=1}} 
  \e^{-\i  { {\vartheta}_{\omega}}(x,y)} \varphi(y).
\end{equation}
The operator $\Delta({ {\vartheta}_{\omega}})$ is bounded (uniformly in
$\omega$), $H({ {\vartheta}_{\omega}},V_{\omega})$ is essentially
self-adjoint on $\cH_{c}$, and  \eqref{lowerbound} holds for some $\gamma \ge 0$.
 \emph{The Anderson Hamiltonian given in \eqref{AND} satisfies these
  assumptions with ${ {\vartheta}_{\omega}}=0$.}

The (random) velocity operator in the $x_{j}$-direction is $\dot{X}_{j} := \i\,
[H,X_{j} ]$, where $X_{j}$ denotes the operator of multiplication by the
$j$-th coordinate $x_{j}$. In the continuum $\dot{X}_{\omega,j}$ is the
closure of the operator $2(-\i \partial_{x_{j}} - \A_{\omega,j})$ defined on
$\mathcal{C}_c^\infty(\RR^d)$, and there is $C_{\gamma}< \infty$ such that
\cite[Proposition~2.3]{BGKS}
\begin{equation}\label{Vbound}
  \bigl\lVert{\dot{X}_{\omega,j}\left(H_\omega +\gamma \right)^{-\frac 1
        2}}\bigr\rVert \le C_{\gamma} \quad \text{for $\PP$-a.e.\  $\omega$}. 
\end{equation}
In the lattice $\dot{X}_{\omega,j}$ is a bounded operator (uniformly in
$\omega$), given by
\begin{equation}
  \begin{split}
    \dot{X}_{\omega,j}&\phantom{:}= D_{j}( {\vartheta}_{\omega}) +\bigl( D_{j}(
      {\vartheta}_{\omega})\bigr)^{*},\\   
    \bigl(D_{j}( {\vartheta}_{\omega})\varphi\bigr)(x)& :=  \e^{-\i  {
        {\vartheta}_{\omega}}(x,x + \widehat{x}_{j})} \varphi(x +
    \widehat{x}_{j})- \varphi(x). 
  \end{split}
\end{equation}

%
\Subsec{The mathematical framework for linear response theory}
The derivation of the Kubo formula will require normed spaces of measurable
covariant operators, which we now briefly describe. We refer to
\cite[Section~3]{BGKS} for background, details, and justifications.

By $\cK_{mc}$ we denote the vector space of measurable covariant operators
$A\colon \Omega \to \mathrm{Lin}\bigl(\cH_{c}, \cH)$, identifying measurable
covariant operators that agree $\PP$-a.e.; all properties stated are assumed
to hold for $\PP$-a.e.\ $\omega \in \Omega$.  Here $\mathrm{Lin}\bigl(\cH_{c},
\cH)$ is the vector space of linear operators from $ \cH_{c}$ to $ \cH$.
Recall that $A$ is measurable if the functions $\omega \to \langle \phi,
A_{\omega} \phi\rangle$ are measurable for all $\phi \in \cH_{c}$, $A$ is
covariant if
\begin{equation}
  U(x)A_{\omega} U(x)^{*}= A_{\tau_{x}(\omega)} \quad \text{for all} \quad x
  \in \ZZ^{d}, 
\end{equation}
and $A$ is locally bounded if $ \|A_\omega \Chi_x\|< \infty$ and $ \|\Chi_x
A_\omega \| <\infty$ for all $ x \in \ZZ^d$.  The subspace of locally bounded
operators is denoted by $\cK_{mc,lb}$.  If $A \in
\cK_{mc,lb}$, we have $\mathcal{D}(A_\omega^*) \supset \cH_{c}$, and hence we may set $A_\omega^\ddagger:=A_\omega^*{\big|}_{\cH_{c}}$.  Note that $(\cJ
A)_{\omega}:= A_\omega^\ddagger$ defines a conjugation in $ \cK_{mc,lb}$.

We introduce norms on $ \cK_{mc,lb}$ given by
\begin{equation}
  \begin{split}
    \tnorm{{A}}_\infty &:=  \|\, \|{A_\omega} \| \,
    \|_{\mathrm{L}^{\infty}(\Omega, \PP)}\\ 
    \tnorm{{A}}_p^{p} &:= \E\bigl\{ \tr \{\Chi_0 |\overline{A_{\omega}}|^{p}
    \Chi_0\}\bigr\}, \quad p=1,2\\ 
    &\phantom{:}= \E\bigl\{\langle \delta_{0}, |\overline{A_{\omega}}|^{p} 
    \delta_{0} \rangle\bigr\} \quad\qquad \text{if} \quad \cH=\ell^{2}(\ZZ^{d}), 
  \end{split}
\end{equation}
and consider the normed spaces
\begin{equation}
  \cK_{p}: = \{ {A}\in \cK_{mc,lb}; \, \tnorm{{A}}_p<\infty \}, \quad
  p=1,2,\infty . 
\end{equation}
It turns out that $\cK_{\infty}$ is a Banach space and $\cK_{2}$ is a Hilbert
space with inner product
\begin{equation}
  \begin{split}
    \llangle A, B\rrangle& :=  \EE \bigl\{ \tr \{\Chi_0
    A_{\omega}^{*}B_{\omega} \Chi_0\} \bigr\}  
    \\
    & \phantom{:}= \EE \bigl\{\langle A_{\omega}\delta_{0}, B_{\omega} 
    \delta_{0}\rangle
    \bigr\}  
    \qquad\quad \text{if} \quad \cH=\ell^{2}(\ZZ^{d}).
  \end{split}
\end{equation}
Since $\cK_{1}$ is not complete, we introduce its (abstract) completion
$\overline{\cK_{1}}$. The conjugation $\cJ$ is an isometry on each $\cK_{p}$,
$ p=1,2,\infty $.  Moreover, $\cK_{p}^{(0)}:= \cK_{p} \cap \cK_{\infty}$ is dense in $\cK_{p}$ for
$p=1,2$.

Note that in the discrete setting we have
\begin{equation} 
  \tnorm{{A}}_1 \le \tnorm{{A}}_2 \le  \tnorm{{A}}_\infty \quad \text{and
    hence} \quad  \cK_{\infty} \subset \cK_{2} \subset \cK_{1};
\end{equation}
in particular, $ \cK_{\infty}= \cK_{p}^{(0)}$ is dense in  $\cK_{p}$, $p=1,2$.
Moreover, in this case we  have $\Delta({ {\vartheta}})$ and
$\dot{X}_{j}$ in $\cK_{\infty}$.

Given $A\in \cK_{\infty}$, we identify $A_{\omega}$ with its closure
$\overline{A_{\omega}}$, a bounded operator in $\cH$.  We may then introduce a
product in $ \cK_{\infty}$ by pointwise operator multiplication, and
$\cK_{\infty}$ becomes a $C^{*}$-algebra. ($\cK_{\infty}$ is actually a von Neumann algebra \cite[Subsection~3.5]{BGKS}.)  This $C^{*}$-algebra acts by left
and right multiplication in $\cK_{p}$, $p=1,2$.  Given $A \in \cK_{p}$, $B \in
\cK_{\infty}$, left multiplication $B \odot_{L} A$ is simply defined by $(B
\odot_{L} A)_{\omega}= B_{\omega} A_{\omega}$.  Right multiplication is more
subtle, we set $(A \odot_{R} B)_{\omega}= A_{\omega}^{\ddagger*} B_{\omega}$
(see \cite[Lemma~3.4]{BGKS} for a justification), and note that $(A \odot_{R}
B)^{\ddagger}= B^{*} \odot_{L}A^{\ddagger}$.  Moreover, left and right
multiplication commute:
\begin{equation}
  B \odot_{L} A \odot_{R} C:= B \odot_{L}( A \odot_{R} C)=(B \odot_{L} A
  )\odot_{R} C 
\end{equation}
for $A \in \cK_{p}$, $B,C \in \cK_{\infty}$.  (We refer to
\cite[Section~3]{BGKS} for an extensive set of rules and properties which
facilitate calculations in these spaces of measurable covariant operators.)

Given $A \in \cK_{p}\, $, $p=1,2$, we define
\begin{align}  \label{U0L}
  \cU^{(0)}_{L}(t) A   & :=\e^{-\i t H} \odot_L  A , \\ 
  \cU^{(0)}_{R}(t) A   & := A \odot_R \e^{- \i t H}, \; \qquad\qquad
  \text{i.e.,}\;\;  \cU^{(0)}_{R}(t)= \cJ \cU^{(0)}_{L}(-t)\cJ \label{U0R},\\ 
  \label{U00}
  \cU^{(0)}(t) A       & :=\e^{-\i t H} \odot_L  A \odot_R 
  \e^{ \i t H}, \; \hspace{.5em} \text{i.e.,} \;\; \cU^{(0)}(t) = 
 \cU^{(0)}_{L}(t)\, \cU^{(0)}_{R}(-t) . 
\end{align}
Then $ \cU^{(0)}(t), \cU^{(0)}_{L}(t), \cU^{(0)}_{R}(t)$ are strongly
continuous one-parameter groups of operators on $\cK_{p}$ for $p=1,2$, which
are unitary on $\cK_2$ and isometric on $\cK_1$, and hence extend to
isometries on $\overline{\cK_1}$. (See \cite[Corollary~4.12]{BGKS} for $
\cU^{(0)}(t)$, the same argument works for $ \cU^{(0)}_{L}(t)$ and $
\cU^{(0)}_{R}(t)$.)  These one-parameter groups of operators commute with each
other, and hence can be simultaneously diagonalized by the spectral theorem.
Using Stone's theorem, we define commuting self-adjoint operators $\cL,
\cH_{L},\cH_{R}$ on $\cK_{2}$ by
\begin{equation}  \label{U00L}
  \e^{-\i t \cL}  := \cU^{(0)}(t), \quad  \e^{-\i t \cH_{L}}  :=
  \cU^{(0)}_{L}(t), 
  \quad  \e^{-\i t \cH_{R}} : = \cU^{(0)}_{R}(t).
\end{equation}
The operator $\cL$ is the \emph{Liouvillian}, we have  
\begin{equation}\label{LHH}
  \cL = \overline{ \cH_{L}- \cH_{R}}  \quad\text{and} \quad   \cH_{R} = \cJ
  \cH_{L} \cJ. 
\end{equation}

 If the ergodic magnetic Schr\"odinger
operator $H $ is bounded, e.g., the Anderson Hamiltonian in \eqref{AND}, then
$H \in \cK_{\infty}$, and $\cL, \cH_{L},\cH_{R}$ are bounded commuting
self-adjoint operators on $\cK_{2}$, with
\begin{equation}
  \cH_{L}A= H\odot_{L }A, \quad  \cH_{R}A= A\odot_{R}H, \quad\text{and} \quad 
  \cL = { \cH_{L}- \cH_{R}}.
\end{equation}

The \emph{trace per unit volume} is given by
\begin{equation}
  \begin{split}
    \cT(A) &:= \E\left\{ \tr\left\{ \Chi_0 A_\omega \Chi_0\right\}\right\}
    \quad\text{for}\quad 
    A \in \cK_1\\
    &\phantom{:}=  \E \bigl\{\langle \delta_{0}, {A}_\omega \delta_{0} 
    \rangle\bigr\} \qquad\quad \text{if} \quad \cH=\ell^{2}(\ZZ^{d}), 
  \end{split}
\end{equation}
a well defined linear functional on $\cK_1$ with $|\cT(A)| \le \tnorm{A}_1$,
and hence can be extended to $\overline{\cK_{1}}$.  Note that $\cT$ is indeed
the {trace per unit volume}:
\begin{equation} \label{tuv}
  \cT(A) = \lim_{L \to \infty}\, \textstyle{ \frac 1 {|\Lambda_L|}}
  \tr\left\{\Chi_{\Lambda_L} A_\omega \Chi_{\Lambda_L}\right\}
  \;\;\;\mbox{for $\PP$-a.e. $\omega$} \, , 
\end{equation}
where $\Lambda_L$ denotes the cube of side $L$ centered at $0$ (see
\cite[Proposition~3.20]{BGKS}).

%
\Subsec{The linear response current}   
We consider a quantum system at zero temperature, modeled by an ergodic
magnetic Schr\"odinger operator $H$ as in \eqref{Homega}.  We fix a Fermi
energy $E_{F}$ and the $x_{1}$-direction, and make the following assumption on
the (random) Fermi projection $P_{E_{F}}:=\Chi_{]-\infty,E_{F}]}(H)$.

\begin{assumption}\label{assumpbes}
\begin{equation}\label{bes}
Y_{{E_{F}}} := \i \left[X_1,P_{E_{F}}\right]\in \cK_2.
\end{equation}
\end{assumption}

Under Assumption~\ref{assumpbes} we have $ Y_{{E_{F}}}=
Y_{{E_{F}}}^{\ddagger}$ and $Y_{E_{F}}\in \mathcal{D}(\cL)$ by
\cite[Lemma~5.4(iii) and Corollary~4.12]{BGKS}.  Moreover, we also have
$Y_{E_{F}} \in \cK_1$ (see \cite[Remark~5.2]{BGKS}).  (Condition \eqref{bes}
is the main assumption in \cite{BGKS}; it was originally identified in
\cite{BESB}.)

\emph{If $H$ is the Anderson Hamiltonian we always have \eqref{bes} if the
  Fermi energy lies in the region of complete localization, i.e., $E_{F}\in
  \Xi^{\mathrm{CL}}$ \cite[Theorem~2]{AiGr98}, \cite[Theorem~3]{GKjsp}.  (In
  fact, in this case $\left[X_j,P_{E_{F}}\right] \in \cK_2$ for all
  $j=1,2,\ldots,d$.)}

In the distant past, taken to be $t=-\infty$, the system is in equilibrium in
the state given by this Fermi projection $P_{E_{F}}$.  A spatially
homogeneous, time-dependent electric field $\mathbf{E}(t)$ is then introduced
adiabatically: Starting at time $t= -\infty$, we switch on the electric field
$\mathbf{E}_{\eta}(t):= \e^{\eta t}\mathbf{E}(t)$ with $\eta >0$, and then let
$\eta \to 0$.\emph{ We here assume that the electric field is pointing in the
  $x_{1}$-direction: $\mathbf{E}(t)=\mathcal{E}(t) \widehat{x}_{1} $, where
  the amplitude $\mathcal{E}(t)$ is a continuous function such  that $ \int_{-\infty}^t \d s \,\e^{\eta s}\abs{\mathcal{E}(s)}
  < \infty$ for all  $t \in \RR$ and $\eta>0$.  Note that the relevant results in \cite{BGKS},
  although stated for constant electric fields $\mathbf{E}$, are valid under
  this assumption.}  
We set $\mathcal{E}_{\eta}(t):= \e^{\eta t}\mathcal{E}(t)$, and
\begin{equation} \label{Ftintro}
  F_{\eta}(t) :=  \int_{-\infty}^t \!\d s \; \mathcal{E}_{\eta}(s).
\end{equation}

For each fixed $\eta>0$ the dynamics are now generated by a time-dependent
ergodic Hamiltonian.  Following \cite[Subsection~2.2]{BGKS}, we resist the
impulse to take $H_{\omega} + \mathcal{E}_{\eta}(t)X_{1}$ as the Hamiltonian,
and instead consider the physically equivalent 
(but bounded below)
Hamiltonian
\begin{equation} 
  H_{\omega}({\eta, t}) := G(\eta,t) H_{\omega}G(\eta,t)^{*},
\end{equation}
where $G(\eta,t):= \mathrm{e}^{\i F_{\eta}(t)X_{1}}$ is a time-dependent gauge
transformation.  We get
\begin{equation} 
  \begin{split} \label{Homegat}
    H_{\omega}({\eta, t})&=H(\A_{\omega}+ F_{\eta}(t)\widehat{x}_{1}
    ,V_{\omega}) \hspace{.85cm} \text{if} \quad \cH=\mathrm{L}^{2}(\RR^{d}),\\ 
    H_{\omega}({\eta, t})&=H({ {\vartheta}_{\omega} }+F_{\eta}(t)
    {\gamma}_{1},V_{\omega}) \hspace{1cm}\text{if} \quad \cH=\ell^{2}(\ZZ^{d}),
  \end{split}
\end{equation}
where $\gamma_{1}(x,y) := y_{1} -x_{1}$ for $(x,y) \in \mathcal{B}(\ZZ^{d})$.

\begin{remark} 
  If $H_{\omega}$ is the Anderson Hamiltonian given in \eqref{AND},
  there is no difficulty in defining $ \widetilde{H}_{\omega}({\eta,
    t}):=H_{\omega} + \mathcal{E}_{\eta}(t)X_{1}$ as a (unbounded)
  self-adjoint operator. Moreover, in this case $ H_{\omega}({\eta, t})$ is
 actually a bounded operator.  It follows that if
  $\widetilde{\psi}(t)$ is a strong solution of the Schr\"odinger
  equation $\i \partial_t \widetilde{\psi}(t) =
  \widetilde{H}_{\omega}({\eta, t}) \widetilde{\psi}(t)$, then
  ${\psi}(t)=G(\eta,t)\widetilde{\psi}(t)$ is a strong solution of $\i
  \partial_t \psi(t) = H_{\omega}({\eta, t}) \psi(t)$.  A similar
  statement holds in the opposite direction for \emph{weak} solutions.
  (See the discussion in \cite[Subsection~2.2]{BGKS}). At the formal level,
 one can easily see that the linear response current  given in  \eqref{lincur2} 
   is independent of the choice of gauge.\end{remark}

The system was described at time $t=-\infty$ by the Fermi projection
$P_{E_{F}}$.  It  is then described at time $t$ by the density matrix
$\varrho(\eta,t)$, the unique solution to the Liouville equation
\eqref{Liouvilleeq} in both spaces $\cK_{2}$ and $\overline{\cK_{1}}$. (See
\cite[Theorem~5.3]{BGKS} for a precise statement.)

The adiabatic electric
field generates a time-dependent electric current. Its amplitude in the
$x_{1}$-direction is given by \eqref{curdef}, where $\dot{X}_{1}(t) :=
G(\eta,t) \dot{X}_{1} G(\eta,t)^{*}$ is the first component of the velocity
operator at time $t$ in the Schr\"odinger picture.  The \emph{linear response
  current} is then defined as in \eqref{lincur2}, its existence  is proven in \cite[Theorem~5.9]{BGKS} with
\begin{align}\label{KuboBGKS}
  J_{\eta,\mathrm{lin}}(t; E_F, \cE) =\cT \left\{ \int_{-\infty}^t \!
    \mathrm{d} r\; \mathrm{e}^{\eta r} \cE(r)
    \dot{X}_{1} \, \cU^{(0)}(t-r)  Y_{E_{F}} \right\}.
    \end{align}
Since the integral in \eqref{KuboBGKS} is a Bochner integral in the Banach
space $\overline{\cK_1}$, where $\cT$ is a bounded linear functional, they can
be interchanged, and hence, using \cite[Eq.~(5.88)]{BGKS}, we obtain
\begin{align}
  J_{\eta,\mathrm{lin}}(t; E_F, \cE)=- \int_{-\infty}^t \! \mathrm{d} r\; \mathrm{e}^{\eta r} \cE(r) \, \llangle
  Y_{E_{F}} , \e^{-\i (t-r)\cL} { \cL} \cP_{E_F} Y_{E_{F}} \rrangle.
  \label{KuboBGKS3}
\end{align}
 Here $\cP_{E_F} $ is the bounded self-adjoint operator on
$\cK_{2}$ given by
\begin{equation}
  \begin{split}\label{cP}
    \cP_{E_F}&:= \Chi_{]-\infty,E_{F}]}(\cH_{L})-
    \Chi_{]-\infty,E_{F}]}(\cH_{R}), \quad \text{that is},\\ 
    \cP_{E_F} A &\phantom{:}=P_{E_{F}} \odot_L A- A  \odot_R P_{E_{F}}
    \quad \mbox{for \ $A \in \cK_{2}$.}
  \end{split}
\end{equation}
Note that $\cP_{E_F}$ commutes with $ { \cL}, \cH_{L},\cH_{R} $; in particular
$\cP_{E_F} {Y_{E_{F}}} \in \mathcal{D}(\cL)$.  Moreover, we have
$\cP_{E_F}^{2} {Y_{E_{F}}}= {Y_{E_{F}}}$ \cite[Lemma~5.13]{BGKS}.

%
\Subsec{The conductivity measure and a Kubo formula for the
  ac-conductivity}  
Suppose now that the amplitude $ \cE(t)$ of the electric field satisfies
assumption \eqref{FT}.  We can then rewrite \eqref{KuboBGKS3}, first using the
Fubini--Tonelli theorem, and then proceeding as in \cite[Eq.~(5.89)]{BGKS}, as
\begin{align}
   \label{Jnu}
   J_{\eta,\mathrm{lin}}(t; E_F, \cE) & = - \int_{\RR}\!\d\nu\;
  \widehat{\cE}(\nu) \int_{-\infty}^{t}\!\d r \; \e^{ (\eta + \i\nu)r} \,
  \llangle 
  {Y_{E_{F}}} ,   \e^{-\i (t-r)\cL} { \cL} \cP_{E_F} {Y_{E_{F}}}\rrangle\\ 
   \qquad  &=-  \i \,\e^{ \eta t} \int_{\RR}\!\d\nu\;  \e^{  \i\nu t}
  \widehat{\cE}(\nu) \,
  \llangle    {Y_{E_{F}}} ,  \left( \cL + \nu - \i\, \eta  \right)^{-1} \left(-
    { \cL}  \cP_{E_F}\right)  {Y_{E_{F}}}\rrangle. \notag
\end{align}
 
This leads us to the following definition, which is justified in the
subsequent theorem.
 
\begin{definition}\label{defSigma} 
  The conductivity measure ($x_{1}$-$x_{1}$ component) at Fermi energy $E_{F}$
  is defined as
  \begin{equation}\label{Sigma}
    \Sigma_{E_{F}}(B):= \pi  \llangle    {Y_{E_{F}}} , \Chi_{B}(\cL) \left(-
      { \cL}  \cP_{E_F}\right)  {Y_{E_{F}}}\rrangle \quad \text{for a Borel
      set  $B \subset\RR$.} 
  \end{equation}
\end{definition}

\begin{theorem}\label{thmSigma} 
  Let $E_{F}$ be a Fermi energy satisfying Assumption~\ref{assumpbes}.  Then
  $\Sigma_{E_{F}}$ is a finite positive even Borel measure on $\RR$.  Moreover, for an electric field with   amplitude $\cE(t)$  satisfying assumption
  \eqref{FT},
we have
   \begin{equation}
  \label{lincur-rig9}
  J_{\eta,\mathrm{lin}}(t; E_F, \cE) = \e^{\eta t} \int_\RR\!\d\nu\;
  \e^{\i\nu t}\, \sigma_{E_{F}}(\eta,\nu) \, \widehat{\cE}(\nu)
\end{equation}
with
    \begin{equation}\label{sigmaetanu}
    \sigma_{E_{F}}(\eta,\nu):=- \tfrac{\i}\pi \int_{\RR} \!\Sigma_{E_{F}}(\d
    \lambda)  \left(\lambda  + \nu - \i \,\eta  \right)^{-1} .
  \end{equation}
  \end{theorem}
  
\Proof 
  Recall that $\cH_{L}$ and $\cH_{R}$ are commuting self-adjoint
  operators on $\cK_{2}$, and hence can be simultaneously diagonalized by the
  spectral theorem. Thus it follows from    \eqref{LHH} and  \eqref{cP} that
  \begin{equation}
    - { \cL}  \cP_{E_F} \ge 0.
  \end{equation}
Since $ {Y_{E_{F}}} \in\mathcal{D}(\cL)$ and $\cP_{E_{F}} $ is bounded, we conclude that  $\Sigma_{E_{F}}$ is a
  finite positive Borel measure.  To show that it is
  even, note that $\mathcal{J}\cL\mathcal{J}= -\cL$, $\mathcal{J} \cP_{E_F}
  \mathcal{J}= - \cP_{E_F} $, and $\mathcal{J}\Chi_{B}(\cL)\cL \cP_{E_F}
  \mathcal{J}=\Chi_{B}(-\cL)\cL \cP_{E_F}= \Chi_{-B}(\cL)\cL \cP_{E_F} $.
  Since $\mathcal{J} {Y_{E_{F}}}= {Y_{E_{F}}}$, we get
  $\Sigma_{E_{F}}(B)=\Sigma_{E_{F}}(-B)$.
  
 Since \eqref{sigmaetanu}  may be rewritten as
  \begin{equation}\label{sigmaetanu1}
  \sigma_{E_{F}}(\eta,\nu)=- \i \,\llangle    {Y_{E_{F}}} ,  \left( \cL + \nu
    -   \i \,\eta  \right)^{-1} \left(- { \cL}  \cP_{E_F}\right)
  {Y_{E_{F}}}\rrangle,
\end{equation}
the equality \eqref{lincur-rig9} follows from \eqref{Jnu}.
 \Endproof

\begin{corollary} \label{corSigma} 
  Let $E_{F}$ be a Fermi energy satisfying Assumption~\ref{assumpbes}, and let
  $\cE(t)$ be the amplitude of an electric field satisfying assumption
  \eqref{FT}.  Then the adiabatic limit $\eta\downarrow 0$ of the linear
  response in phase current given in \eqref{lincur-rig-in} exists:
  \begin{align}
    \label{in-rig25}
    J_{\mathrm{lin}}^{\mathrm{in}}(t; E_F, \cE):=  \lim_{\eta\downarrow
      0}J_{\eta,\mathrm{lin}}^{\mathrm{in}}(t; E_F,  \cE) =
    \int_\RR\!\Sigma_{E_{F}}(\d\nu)\; \e^{\i\nu t}\, 
    \widehat{\cE}(\nu).
  \end{align}
  If in addition $\cE(t)$ is uniformly H\"older continuous, then the
  adiabatic limit $\eta\downarrow 0$ of the linear response out of
  phase current also exists:
  \begin{equation}
    \label{out-rig}
    \begin{split}
    J_{\mathrm{lin}}^{\mathrm{out}}(t; E_F, \cE):&=  \lim_{\eta\downarrow 0}
    J_{\eta,\mathrm{lin}}^{\mathrm{out}}(t; E_F,  \cE) \\
    & = \tfrac 1 {\pi \i}
    \int_\RR\!\Sigma_{E_{F}}(\d\lambda)\;  \; \mbox{\textsc{pv}} \!\! 
    \int_\RR \!\d\nu\; \frac{\e^{\i\nu t } \,
      \widehat{\cE}(\nu)}{\nu-\lambda},
  \end{split}
  \end{equation}
  where the integral over $\nu$ in \eqref{out-rig} is to be understood in the
  principal-value sense.
\end{corollary}

\Proof  This corollary is an immediate consequence of \eqref{lincur-rig9},
\eqref{sigmaetanu}, and well known properties of the Cauchy  (Borel, Stieltjes) transform of finite Borel measures.  The limit in  \eqref{in-rig25} follows from \cite[Theorem~2.3]{StWe71}. The limit in \eqref{out-rig} can be established using Fubini's theorem and the existence (with bounds) of the principal value integral for uniformly H\"older continuous functions (see \cite[Remark~4.1.2]{Gra04}).
\Endproof
  
\begin{remark} 
  The out of phase (or passive) conductivity does not appear to be the subject
  of extensive study;   but see \cite{LiGr88}.
\end{remark}

%
\Subsec{Correlation measures}
For each $A \in \cK_{2}$ we  define a finite Borel measure $\Upsilon_{A}$ on
$\RR^{2}$ by
\begin{equation}\label{Psi}
  {\Upsilon_{A}}(C) := \llangle  A, \Chi_{C}(\cH_{L}, \cH_{R})  A \rrangle  \quad
  \text{for a Borel set $C \subset\RR^{2}$}. 
\end{equation}
Note that it follows from \eqref{LHH} that 
\begin{equation}
\Upsilon_{A}(B_{1}\times B_{2})=\Upsilon_{A^{\ddagger}}(B_{2}\times B_{1})\quad \text{for all Borel sets $B_{1},B_{2} \subset \RR$}.
\end{equation}

The correlation measure we obtain by taking $A=Y_{E_{F}}$ plays an important role in our analysis.

\begin{proposition} \label{SigmaUpsilon5}
Let $E_{F}$ be a Fermi energy satisfying Assumption~\ref{assumpbes} and set   $\Psi_{E_{F}}:= \Upsilon_{Y_{E_{F}}}$.  Then
 \begin{equation}\label{SigmaUpsilon}
     \Sigma_{E_{F}}(B)=
     \pi  \int_{\RR^{2}} \!{\Psi_{E_{F}}}(\d\lambda_{1}\d\lambda_{2})
     \; \abs{\lambda_{1} -\lambda_{2}}\,
     \Chi_{B}(\lambda_{1} - \lambda_{2}) 
   \end{equation}
   for all Borel sets $B \subset \RR$. Moreover, the measure  $  {\Psi_{E_{F}}}$ is  supported by the set $ \mathbb{S}_{E_{F}}$, i.e., $ {\Psi_{E_{F}}}(\RR^{2}\setminus\mathbb{S}_{E_{F}})=0 $, where
 \begin{equation}
     \mathbb{S}_{E_{F}}:= \bigl\{]-\infty,E_{F}] \times ]E_{F},\infty[\bigr\}
     \cup \bigl\{ ]E_{F},\infty[\times ]-\infty,E_{F}]\bigr\} \subset
     \RR^{2}. 
   \end{equation}
\end{proposition} 
  
\Proof 
If we set
  \begin{equation}
    Q_{E_{F}}(\lambda_{1},\lambda_{2}):=\Chi_{\mathbb{S}_{E_{F}}}(
    \lambda_{1}, \lambda_{2})      
    = \bigl|\Chi_{]-\infty,E_{F}]}(\lambda_{1}) - 
      \Chi_{]-\infty,E_{F}]}(\lambda_{2})\bigr|,
  \end{equation}
  it follows from \eqref{cP} that
  \begin{equation}
    \cQ_{E_{F}}=\cP_{E_{F}}^{2}, \quad \text{where} \quad
    \cQ_{E_{F}}:=Q_{E_{F}}(\cH_{L},\cH_{R}). 
  \end{equation}
  Thus $ \cQ_{E_{F}} {Y_{E_{F}}}= {Y_{E_{F}}}$, and 
 the measure $ {\Psi_{E_{F}}}$ is
  supported by the set  $ \mathbb{S}_{E_{F}}$.
Hence
    \begin{equation}\label{Sigma8}
    \Sigma_{E_{F}}(B)= \pi  \llangle    {Y_{E_{F}}} , \Chi_{B}(\cL) 
     \abs{ \cL}   {Y_{E_{F}}} \rrangle \quad \text{for all Borel
      sets $B \subset\RR$}, 
  \end{equation}
and \eqref{SigmaUpsilon} follows.
\Endproof

%
\Subsec{The velocity-velocity correlation measure}
The velocity-velocity correlation measure $\Phi$ is formally given by $\Phi =
\Upsilon_{\dot{X}_{1}}$, but note that $\dot{X}_{1}\notin \cK_{2}$ in the
continuum setting.

\begin{definition}\label{defPhi}  
  The velocity-velocity correlation measure ($x_{1}$-$x_{1}$ component) is the
  positive $\sigma$-finite Borel measure on $\RR^{2}$ defined on bounded Borel
  sets $C \subset \RR^{2}$ by
  \begin{align}\label{Vchi}
    \Phi(C) &:=   \llangle  {\dot{X}_{1,\alpha}},   \left(\cH_{L} +\gamma
    \right)^{2\alpha}\Chi_{C}(\cH_{L}, \cH_{R})  
    \left(\cH_{R} +\gamma \right)^{2\alpha} {\dot{X}_{1,\alpha}}\rrangle \\
     & \phantom{:}= \Upsilon_{\dot{X}_{1}}(C)  \quad \text{if} \quad
    \cH=\ell^{2}(\ZZ^{d}),\label{Vchi2} 
  \end{align}
  where
  \begin{equation}\label{tildeV}
    \begin{split}
      {\dot{X}_{1,\alpha}}&:=\left\{\left(H +\gamma
        \right)^{-\alpha}\dot{X}_{1} 
        \left(H +\gamma \right)^{-\frac 1 2}\right\}\odot_{L} \left(H +\gamma
      \right)^{- [[\frac d 4 ]]} \in \cK_{2},\\
      \alpha& :=\tfrac 1 2 + [[\tfrac d 4 ]] \quad \text{with $[[\tfrac d 4
        ]]$ the smallest integer bigger than $\tfrac d 4 $}.
      \end{split}
    \end{equation}
\end{definition}

Note that \eqref{tildeV} is justified since we have  $\dot{X}_{1} \left(H +\gamma
\right)^{-\frac 1 2} \in \cK_{\infty}$ by \eqref{Vbound} and $ \left(H +\gamma
\right)^{- [[\frac d 4 ]]} \in \cK_{2}$ by \cite[Proposition~4.2(i)]{BGKS}; note ${\dot{X}_{1,\alpha}}^{\ddagger}={\dot{X}_{1,\alpha}}$.  In the
discrete setting we have $\dot{X}_{1}\in \cK_{2}$ and hence $\Phi =
\Upsilon_{\dot{X}_{1}}$, a finite measure .

The following lemma relates the measure ${\Psi_{E_{F}}}$  of Proposition~\ref{SigmaUpsilon5} to the measure $\Phi$.

\begin{lemma} The correlation measure ${\Psi_{E_{F}}}$ is absolutely continuous with respect to the  velocity-velocity correlation measure $\Phi$, with
  \begin{equation}\label{UpsilonPhi}
 \frac{  { \d \Psi_{E_{F}}}} { \d \Phi}    (\lambda_{1},\lambda_{2})= 
   \frac { Q_{E_{F}} (\lambda_{1},\lambda_{2})}{ (\lambda_{1}-\lambda_{2})^{2}}. 
  \end{equation}
\end{lemma}

\Proof
  The key observation is that (use \cite[Lemma~5.4(iii) and
  Corollary~4.12]{BGKS})
  \begin{equation}
    \begin{split}
      \left(\cH_{L} +\gamma \right)^{-\alpha} \left(\cH_{R} +\gamma
      \right)^{-\alpha} 
      \cL  Y_{E_{F}} &= -\cP_{E_{F}} {\dot{X}_{1,\alpha}},\\
      \cL  Y_{E_{F}} &= -\cP_{E_{F}} {\dot{X}_{1}}  \quad \text{if} \quad
      \cH=\ell^{2}(\ZZ^{d}).
    \end{split}
  \end{equation}
  It follows that for all Borel sets $C\subset \RR^{2}$ we have
  \begin{equation}
    \begin{split}
      &\int_{C} \! {\Psi_{E_{F}}}(\d\lambda_{1}\d\lambda_{2}) \;
      (\lambda_{1}-\lambda_{2})^{2} = \llangle \cL
      {Y_{E_{F}}},\Chi_{C}(\cH_{L}, \cH_{R}) \cL   {Y_{E_{F}}}\rrangle\\
      &\qquad = \llangle\cP_{E_{F}} {\dot{X}_{1,\alpha}}, \left(\cH_{L}
        +\gamma \right)^{2\alpha}\Chi_{C}(\cH_{L}, \cH_{R}) \left(\cH_{R}
        +\gamma \right)^{2\alpha}\cP_{E_{F}}
      {\dot{X}_{1,\alpha}}\rrangle\\
      &\qquad = \llangle {\dot{X}_{1,\alpha}}, \cP_{E_{F}}^{2}
      \left(\cH_{L} +\gamma \right)^{2\alpha}\Chi_{C}(\cH_{L}, \cH_{R})
      \left(\cH_{R} +\gamma \right)^{2\alpha} {\dot{X}_{1,\alpha}}\rrangle\\
      &\qquad = \llangle {\dot{X}_{1,\alpha}}, \left(\cH_{L} +\gamma
      \right)^{2\alpha}\Chi_{C\cap \mathbb{S}_{E_{F}}}(\cH_{L}, \cH_{R})
      \left(\cH_{R} +\gamma \right)^{2\alpha} {\dot{X}_{1,\alpha}}\rrangle\\
      &\qquad = \int_{C}\!  \Phi(\d\lambda_{1} \d\lambda_{2}) \; Q_{E_{F}}
      (\lambda_{1},\lambda_{2}) . 
    \end{split}
  \end{equation}
 Since $ {\Psi_{E_{F}}}$ is supported on $\mathbb{S}_{E_{F}}$, the lemma follows.
\Endproof

We can now write the conductivity measure in terms of the  velocity-velocity
correlation measure.

\begin{proposition}  \label{PropSigmaPhi} 
  Let $E_{F}$ be a Fermi energy satisfying Assumption~\ref{assumpbes}.  Then
  \begin{equation} \label{SigmaPhi}
    \Sigma_{E_{F}}(B)=
    \pi  \int_{\Sbb_{E_{F}}}  \! \Phi(\d\lambda_{1} \d\lambda_{2})
    \; \abs{\lambda_{1} -\lambda_{2}}^{-1}
    \Chi_{B}(\lambda_{1} - \lambda_{2}) 
  \end{equation}
  for all Borel sets $B \subset \RR$.
\end{proposition}

\Proof
  The representation \eqref{SigmaPhi} is an immediate consequence of
  \eqref{SigmaUpsilon} and \eqref{UpsilonPhi}.
\Endproof

\begin{remark}  
  If we assume, as customary in physics, that the conductivity
  measure $\Sigma_{E_{F}}$ is absolutely continuous, its density being the in
  phase conductivity $ \sigma_{E_{F}}^{\mathrm{in}}(\nu)$, and that in addition the
  velocity-velocity correlation measure $ \Phi$ is absolutely continuous with
  a continuous density $\phi(\lambda_{1},\lambda_{2})$, then \eqref{SigmaPhi}
  yields the well-known formula (cf.\ \cite{Pas99,KiLe03})
  \begin{equation}
    \label{physKubo}
    \sigma_{E_{F}}^{\mathrm{in}}(\nu) = \frac{\pi}{\nu}
    \int_{E_{F}-\nu}^{E_{F}} \!\d 
    E\;  \phi( E+\nu,E).
  \end{equation}
  The existence of the densities $ \sigma_{E_{F}}^{\mathrm{in}}(\nu) $ and
  $\phi(\lambda_{1},\lambda_{2})$ is currently an open question, and hence
  \eqref{physKubo} is only known as a formal expression. In contrast, the
  integrated version \eqref{SigmaPhi} is mathematically well established. (See
  also \cite{BH} for some recent work on the velocity-velocity correlation
  function.)
\end{remark}

%
\Subsec{Bounds on the average in phase conductivity}
The average in phase conductivity
$\overline{\sigma}_{E_{F}}^{\mathrm{in}}(\nu)$ defined in
\eqref{overlinesigma} can be bounded from above and below by the correlation measure $\Psi_{E_{F}}$.
Note that since $\Sigma_{E_{F}}$ is an even measure it suffices to consider
frequencies $\nu > 0$.

\begin{proposition}\label{sigmaUpsilon} 
  Let $E_{F}$ be a Fermi energy satisfying Assumption~\ref{assumpbes}.  Given
  $\nu>0$, define the pairs of disjoint energy intervals
  \begin{equation}
    \begin{split}\label{I+I-}
      I_- :=  ]E_{F} - \nu, E_{F}]\quad& \text{and} \quad I_+ := ]E_{F}, E_{F}
      +\nu], \\ 
      J_- :=    ]E_{F} - \tfrac \nu 2, E_{F} -\tfrac \nu 4]
      \quad& \text{and} \quad J_+ :=  ]E_{F} + \tfrac \nu 4, E_{F} + \tfrac
      \nu 2]. 
    \end{split}
  \end{equation}
  Then
  \begin{equation} \label{overlinesigma2}
    \tfrac \pi  2\, \Psi_{E_{F}}(J_{+ }\times J_{-}) \le
    \overline{\sigma}_{E_{F}}^{\mathrm{in}}(\nu) \le \pi\,
    \Psi_{E_{F}}(I_{+ }\times I_{-}). 
  \end{equation}
\end{proposition}

\begin{figure}[t]
  \begin{center}
    \leavevmode
    \includegraphics[width=0.6\textwidth]{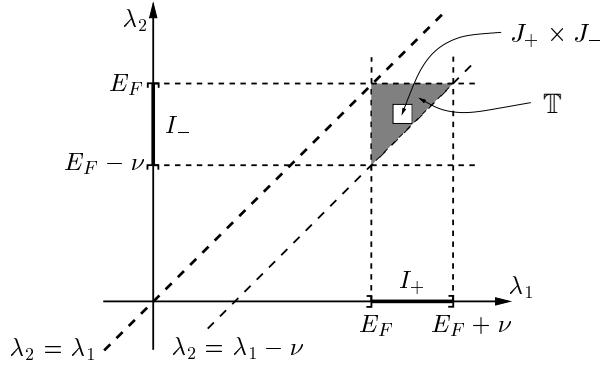}
    \caption{\mbox{\label{fig:triangle}}%
      $J_{+ }\times J_{-}\subset \mathbb{T}\subset I_{+}\times I_{-}$.}
  \end{center}
\end{figure}

\Proof
  It follows immediately from the representation \eqref{SigmaUpsilon} that
  \begin{equation}
    \label{overlinesigma3}
    \overline{\sigma}_{E_{F}}^{\mathrm{in}}(\nu) \le 
    \pi  \int_{\Sbb_{E_{F}}} \!\!{\Psi_{E_{F}}}(\d\lambda_{1}
    \d\lambda_{2})\; 
    \Chi_{[0,\nu]}(\lambda_{1} - \lambda_{2})= \pi  \,
    \Psi_{E_{F}}(\mathbb{T}), 
  \end{equation}
  where
  \begin{equation}
    \mathbb{T}:= \{ (\lambda_{1},\lambda_{2}) \in \mathbb{R}^{2} : \lambda_{2} 
    \le E_{F} < \lambda_{1} \text{~~and~~} \lambda_{1} - \lambda_{2} \le \nu\} 
  \end{equation}
  is the triangle in Figure~\ref{fig:triangle}.  Since
  $\mathbb{T}\subset I_{+ }\times I_{-}$, as can be seen in
  Figure~\ref{fig:triangle}, the upper bound in \eqref{overlinesigma2} follows
  from \eqref{overlinesigma3}.
  
  Similarly, we have $J_{+ }\times J_{-} \subset \mathbb{T} $ (see
  Figure~\ref{fig:triangle}) and the lower bound in \eqref{overlinesigma2}.
\Endproof

%
\section{The proof  of Theorem \ref{ourmott}}
\label{sec:mottproof}
\emph{In this section we let $H$ be  the Anderson Hamiltonian and fix a Fermi
  energy $E_{F} \in \Xi^{\mathrm{CL}}$.}  Thus \eqref{aizenmolch} holds, and
hence, using the exponential decay of the Fermi projection given in
\cite[Theorem~2]{AiGr98} and $\norm{P_{E_{F}} }\le 1$, we have
\begin{equation}\label{Pdecay}
  \E \left \{ \left\lvert\langle\delta_{x}, P_{E_{F}}
      \delta_{y}\rangle\right\rvert^{p}\right\} 
  \le C \e^{-c \abs{x-y}} \quad  \quad \text{for all $p\in [1,\infty[ $ and $
    x,y \in \ZZ^{d}$},
\end{equation} 
where $C$ and $c >0$ are constants depending on $E_{F}$ and $\rho$.  In
particular, Assumption~\ref{assumpbes} is satisfied, and we can use the
results of Section~\ref{sec:Kubo}.

In view of Proposition~\ref{sigmaUpsilon}, Theorem \ref{ourmott} is an
immediate consequence of the following result.

\begin{theorem}
  \label{ourmott2}
  Let $H$ be the Anderson Hamiltonian and consider a Fermi energy in its
  region of complete localization: $E_{F} \in \Xi^{\mathrm{CL}}$. Consider the
  finite Borel measure $ \Psi_{E_{F}}$ on $ \RR^{2}$ of Proposition~\ref{SigmaUpsilon5},
   and, given $\nu>0$, let $I_{-}$ and $I_{+}$ be the disjoint
  energy intervals given in \eqref{I+I-}.  Then
  \begin{equation}
    \label{leadingboundN}
    \limsup_{\nu \downarrow 0} \frac { \Psi_{E_{F}}(I_{+}\times I_{-}) } 
    { \nu^{2} \left(\log\tfrac{1}{\nu}\right)^{d+2}}\le  {205}^{d+2}
    { \pi^2}  \, \|\rho\|_\infty^2
    {\ell}_{E_{F}}^{d+2},
  \end{equation}
  where $\ell_{E_{F}}$ is as in \eqref{aizenmolch} and $\rho$ is the density
  of the single site potential.
\end{theorem}

Theorem~\ref{ourmott2} will be proven by a reduction to finite volume, a cube
of side $L$, where the relevant quantity will be controlled by Minami's
estimate.  Optimizing the final estimate will lead to a choice of $L \sim
\log\frac{1}{\nu}$, which is responsible for the factor of $
\left(\log\tfrac{1}{\nu}\right)^{d+2}$ in \eqref{leadingboundN}. By improving
some of the estimates in the proof (at the price of making them more
cumbersome), the numerical constant $205$ in \eqref{leadingboundN} may be
reduced to $36$.

%
\Subsec {Some properties of the measure $ \Psi_{E_{F}}$}
We briefly recall some facts about the Anderson Hamiltonian. If $I \subset
\Xi^{\mathrm{CL}}$ is a compact interval, then for all Borel functions $f$
with $\abs{f}\le1$ we have \cite{Aiz94,AiGr98}
\begin{equation}\label{PdecayI}
  \E \left \{ \left\lvert\langle\delta_{x},f(H) \Chi_{I}(H)
      \delta_{y}\rangle\right\rvert\right\} 
  \le C_{I}\, \e^{-c_{I} \abs{x-y}} \quad  \quad \text{for all $ x,y \in
    \ZZ^{d}$}, 
\end{equation}
for suitable constants $C_{I}$ and $c_{I} >0$, and hence
\begin{equation}\label{loccom}
  \left[X_1,f(H) \Chi_{I}(H)\right]\in \cK_2.
\end{equation}
We also recall Wegner's estimate \cite{Weg81}, which  yields
\begin{equation} \label{wegnerest}
  \left\lvert\E \left \{\langle\delta_{x},
      \Chi_{B}(H)\delta_{y}\rangle\right\} \right\rvert\le 
  \E \left \{\langle\delta_{0}, \Chi_{B}(H)\delta_{0}\rangle\right\} \le
  \norm{\rho}_{\infty}\abs{B} 
\end{equation}
for all Borel sets $B\subset\RR$ and $x,y \in \ZZ^{d}$.

We begin by proving a preliminary bound on $ \Psi_{E_{F}}(I_{+}\times
I_{-})$, a consequence of Wegner's estimate.

\begin{lemma}
  \label{wegner}
  Given $\beta \in ]0,1[$,
  there exists a  constant $W_{\beta} $ such that
  \begin{equation}
    \Psi_{E_{F}}(B_{+}\times B_{-}) 
    \le W_{\beta}\left( \min\bigl\{ |B_{+}|, |B_{-}|\bigr\} \right)^{\beta}
  \end{equation}
  for all  Borel sets $B_{\pm}\subset \mathbb{R}$.
\end{lemma}

\Proof     
  Since
  \begin{align}
    \Psi_{E_{F}}(B_{+ }\times B_{-}) &\le \min \left\{
      \Psi_{E_{F}}(B_{+ }\times \RR), 
      \Psi_{E_{F}}(\RR \times B_{-})\right\},\\
    \Psi_{E_{F}}(B_{+ }\times B_{-})& =  \Psi_{E_{F}}(B_{-}\times
    B_{+}),\\ 
    \intertext{and, for all Borel sets $B \subset \RR$,}
    \Psi_{E_{F}}(B\times \RR)&=   \llangle Y_{E_{F}},
    \Chi_{B}(\mathcal{H}_{L}) Y_{E_{F}} \rrangle, 
  \end{align}
  it suffices to show that for $\beta \in ]0,1[$ there exists a constant
  $W_{\beta} $ such that
  \begin{equation}
    \llangle Y_{E_{F}}, \Chi_{B}(\mathcal{H}_{L}) Y_{E_{F}} \rrangle \le
    W_{\beta} \abs{B}^{\beta} \quad \text{for all Borel sets $B\subset\RR$}. 
  \end{equation}
 Using  $X_{1} \delta_{0}=0$, we obtain
  \begin{align}
    \label{hoelderchain}
     \hspace*{1cm} \llangle Y_{E_{F}}&, \Chi_{B} (\mathcal{H}_{L}) Y_{E_{F}}
    \rrangle 
    = \EE \bigl\{ \langle X_{1} P_{E_{F}} \delta_{0}, \Chi_{B}(H) X_{1}
    P_{E_{F}} 
    \delta_{0} \rangle\bigr\} \nonumber\\
    & \le \sum_{x,y \in \ZZ^{d}} |x_{1}| |y_{1}| \;\EE \bigl\{ 
    |\langle\delta_{0}, P_{E_{F}}\delta_{x}\rangle| \,
    |\langle\delta_{x}, \Chi_{B}(H) \delta_{y}\rangle| \,
    |\langle\delta_{y}, P_{E_{F}}\delta_{0}\rangle| \bigr\} \\
    & \le W_{\beta}\abs{B}^{\beta}, \nonumber
  \end{align}
  where we used H\"older's inequality plus the estimates \eqref{Pdecay} and
  \eqref{wegnerest}.
\Endproof

\begin{remark} 
  In the case of the Anderson Hamiltonian, the self-adjoint operators
  $\cH_{L}$ and $\cH_{R}$ on the Hilbert space $\cK_{2} $ have absolutely
  continuous spectrum.  The proof is a variation of the argument in
  Lemma~\ref{wegner}. Recalling that in the discrete setting $\cK_{\infty} $
  is a dense subset of $\cK_{2}$, to show that $\cH_{L}$ has absolutely
  continuous spectrum it suffices to prove that for each $A \in \cK_{\infty}$
  the measure $\Upsilon_{A}^{(L)}$ on $\RR$, given by $ \Upsilon_{A}^{(L)}(B)
  := \Upsilon_{A}(B\times \RR)$ (see \eqref{Psi}) is absolutely continuous.
  Since $ \Chi_{B}(H)\in \cK_{\infty}\subset \cK_{2} $, we have, similarly to
  \eqref{hoelderchain}, that
  \begin{align}
    \hspace*{1cm}\Upsilon_{A}^{(L)}(B) & =   
       \llangle A, \Chi_{B}(\mathcal{H}_{L}) A \rrangle 
    =\tnorm{ \Chi_{B}(H)\odot_{L} A }_{2}^{2}=\tnorm{\smash{A^{\ddagger}}
    \odot_{R} 
      \Chi_{B}(H) }_{2}^{2} 
    \nonumber\\
    &  \le \tnorm{A}_{\infty}^{2} \tnorm{  \Chi_{B}(H) }_{2}^{2}=
    \tnorm{A}_{\infty}^{2}\E \left \{\langle\delta_{0},
      \Chi_{B}(H)\delta_{0}\rangle\right\}\\ 
    &  \le  \norm{\rho}_{\infty} \tnorm{A}_{\infty}^{2} \abs{B}.\nonumber 
  \end{align}
  Unfortunately, knowing that $\cH_{L}$, and hence also $\cH_{R}$, has
  absolutely continuous spectrum does not imply that the Liouvillian $\cL=
  \cH_{L}-\cH_{R}$ has no nonzero eigenvalues.  (Note that $0$ is always an
  eigenvalue for $\cL$.)
\end{remark}

The next lemma rewrites $ \Psi_{E_{F}}(I_{+}\times I_{-})$ in ordinary
$\ell^{2}(\ZZ^{d})$-language. Recall that $f(H)\in \cK_{2}\cap \cK_{\infty}$
and $ [X_{1},f(H)] \in \cK_{2}$ if either $f \in \mathcal{S}(\RR)$, or $f$ is
a bounded Borel function with $f \Chi_{I}=f$ for some bounded interval $I
\subset \Xi^{\mathrm{CL}}$, or $f=\Chi_{]-\infty,E]}$ with $E \in
\Xi^{\mathrm{CL}}$ \cite[Proposition 4.2]{BGKS}.

\begin{lemma}
  \label{dictionary} 
  Let $F_{\pm}:=f_{\pm}(H)$, where $f_{\pm}\ge 0$ are bounded Borel measurable
  functions on $\RR$.  Suppose
  \begin{equation}\label{FF}
    F_{-}P_{E_{F}} = F_{-},  \quad F_{+}P_{E_{F}}=0, \quad \text{and} \quad F_{\pm},
    [X_{1},F_{\pm}]   \in \cK_{2}. 
  \end{equation}
  Then 
  \begin{equation}\label{XFXF}
    \int_{\RR^{2}}  \!\Psi_{E_{F}}(\d\lambda_{1} \d\lambda_{2}) \;
    f_{+}^{2}(\lambda_{1}) f_{-}^{2}(\lambda_{2} )= 
    \EE\bigl\{ \langle\delta_{0}, F_{-} X_{1}F_{+}^{2} X_{1}
    F_{-} \delta_{0} \rangle\bigr\} \,. 
  \end{equation}
\end{lemma}

\Proof  
  It follows from  \eqref{Psi} that 
  \begin{equation}\label{FodotF}
    \int_{\RR^{2}} \! \Psi_{E_{F}}(\d\lambda_{1} \d\lambda_{2}) \;
    f_{+}^{2}(\lambda_{1}) f_{-}^{2}(\lambda_{2} ) = \tnorm{F_{+}\odot_{L}
      Y_{E_{F}} \odot_{R} F_{-}}_{2}^{2}. 
  \end{equation}
  In view of \eqref{bes} and \eqref{FF}, it follows from
  \cite[Eq.~(4.8)]{BGKS} that
  \begin{equation}
    -\i Y_{E_{F}} \odot_{R} F_{-}=  [X_{1}, F_{-}P_{E_{F}}] -
    P_{E_{F}}\odot_{L}  [X_{1}, F_{-}]  
    = [X_{1}, F_{-}] - P_{E_{F}}\odot_{L}  [X_{1}, F_{-}],
  \end{equation}
  and hence
  \begin{equation}
    F_{+}\odot_{L} Y_{E_{F}} \odot_{R} F_{-}= \i\,  F_{+}\odot_{L}[X_{1},
    F_{-}] . 
  \end{equation}
  Thus it follows from \eqref{FodotF} that 
  \begin{equation}
    \int_{\RR^{2}} \!\Psi_{E_{F}}(\d\lambda_{1} \d\lambda_{2}) \;
    f_{+}^{2}(\lambda_{1}) f_{-}^{2}(\lambda_{2} ) =\E\left\{  \norm{F_{+}X_{1}
        F_{-}\delta_{0}}_{2}^{2}\right\}, 
  \end{equation}
  which implies  \eqref{XFXF}.
\Endproof  
  
Lemma~\ref{dictionary} has the following corollary, which will be used to
justify the replacement of spectral projections by smooth functions of $H$.

\begin{lemma}
  \label{dictionary2} 
  Let $B_{\pm}$ be bounded Borel subsets of the region of complete
  localization $\Xi^{\mathrm{CL}}$ with $B_{-} \subset ]-\infty, E_{F}]$ and
  $B_{+}\cap ]-\infty,E_{F}] = \emptyset$, so
  \begin{equation}\label{P+P-}
    P_{-}P_{E_{F}} = P_{-} \quad \text{and} \quad P_{+}P_{E_{F}}=0, \quad
    \text{where}\quad P_{\pm} := \Chi_{B_{\pm}}(H), 
  \end{equation}
  and let $f_{\pm}$ and $F_{\pm}$ be as in Lemma~\ref{dictionary} obeying 
  $\Chi_{B_{\pm}}\le f_{\pm} \le 1$.  Then
  \begin{align}\label{XPXP}
    \Psi_{E_{F}}(B_{+}\times B_{-})& =
    \EE\bigl\{ \langle\delta_{0}, P_{-} X_{1}P_{+} X_{1}
    P_{-} \delta_{0} \rangle\bigr\}\\
    &\le   \EE\bigl\{ \langle\delta_{0}, F_{-} X_{1}F_{+} X_{1}
    F_{-} \delta_{0} \rangle\bigr\} .\label{XPXP2}
  \end{align}
\end{lemma}

\Proof  
  The equality \eqref{XPXP} follows from Lemma~\ref{dictionary} with $f_{\pm}=
  \Chi_{B_{\pm}}$.  To prove the bound \eqref{XPXP2}, note that we also have $
  \Chi_{B_{\pm}}\le f_{\pm}^{2} \le f_{\pm}\le 1$, and hence, since
  \begin{equation}
    \Psi_{E_{F}}(B_{+}\times B_{-})
    \le  \int_{\RR^{2}} \!\Psi_{E_{F}}(\d\lambda_{1}\d\lambda_{2}) \;
    f_{+}^{2}(\lambda_{1}) f_{-}^{2}(\lambda_{2}), 
  \end{equation}
  \eqref{XPXP2} follows from  \eqref{XFXF} since $F_{+}^{2}\le F_{+}$. 
\Endproof

%
\Subsec{Passage to finite volume}
\label{secondpart}
Restricting the Anderson Hamiltonian to finite volume leads to a natural
minimal distance between its eigenvalues, as shown in \cite[Lemma~2]{KlMo05}
using Minami's estimate \cite{Min96}.  It is this natural distance that allows
control over an eigenvalue correlation like \eqref{XPXP}.

The finite volumes will be cubes $\Lambda_{L}$ with $L\ge 3$.  Here
$\Lambda_{L}$ is the largest cube in $\ZZ^d$, centered at the origin and
oriented along the coordinate axes, with $\abs{\Lambda_{L}} \le L^d$.  We
denote by $H_{L}$ the (random) \emph{finite-volume restriction} of the
Anderson Hamiltonian $H$ to $\ell^{2}(\Lambda_{L})$ \emph{with periodic
  boundary condition}. We will think of $\ell^{2}(\Lambda_{L})$ as being
naturally embedded into $\ell^{2}(\ZZ^{d})$, with all operators defined on
$\ell^{2}(\Lambda_{L})$ acting on $\ell^{2}(\ZZ^{d})$ via their trivial
extension.  In addition, it will be convenient to consider another extension
of $H_{L}$ to $\ell^{2}(\ZZ^{d})$, namely
\begin{equation}
  \widehat{H}_{L} := H_{L} + \Chi_{\Lambda_{L}^{c}}H \Chi_{\Lambda_{L}^{c}},
\end{equation}
where by $S^{c}$ we denote the complement of the set $S$.  We set $\partial S
:= \{ x\in S\colon \, \text{there exists $y\in S^{c}$ with $|x-y| =1$}\}$, the
boundary of a subset $S$ in $\ZZ^{d}$.  Moreover, when convenient we use the
notation $A(x,y):= \langle \delta_x, A \delta_y \rangle$ for the matrix
elements of a bounded operator $A$ on $\ell^{2}(\ZZ^{d})$.

To prove \eqref{leadingboundN}, we rewrite $ \Psi_{E_{F}}(I_{+}\times I_{-})$
as in \eqref{XPXP}, estimate the corresponding finite-volume quantity, and
calculate the error committed in going from infinite to finite volume. To do
so, we would like to express the spectral projections in \eqref{XPXP} in terms
of resolvents, where we can control the error by the resolvent identity.  This
can be done by means of the Helffer--Sj\"ostrand formula for \emph{smooth}
functions $f$ of self-adjoint operators \cite{HeSj89,HuSi00}.  More precisely,
it requires finiteness in one of the norms
\begin{equation} \label{sdfn}
  \hnorm{f}_m := \sum_{r=0}^m \int_{\mathbb{R}}\!\mathrm{d}u\;
  |f^{(r)}(u)|\,(1 + \abs{u}^{2})^{\frac {r-1} 2}  , \quad  m=1,2,\ldots \,. 
\end{equation}
If $ \hnorm{f}_m < \infty$ with $m \ge 2$, then for any self-adjoint operator
$K$ we have
\begin{equation}\label{HS}
  f (K) = \int_{\RR^{2}} \!\d\tilde{f}(z) \, (K-z)^{-1} ,
\end{equation}
where the integral converges absolutely in operator norm.  Here $z= x + \i y$,
$\tilde{f}(z)$ is an \emph{almost analytic extension} of $f$ to the complex
plane, $\d\tilde{f}(z) := \frac 1 {2\pi}\partial_{\bar{z}}\tilde{f}(z)
\,\mathrm{d} x\, \mathrm{d} y $, with $\partial_{\bar{z}}= \partial_x + \i
\partial_y$, and $|\d\tilde{f}(z)| := (2\pi)^{-1}
|\partial_{\,\overline{z}}\tilde{f}(z)| \,\mathrm{d} x\, \mathrm{d} y$.
Moreover, for all $p \ge 0$ we have
\begin{equation}\label{HShigherorder}
  \int_{\RR^{2}} \! |\d\tilde{f}(z)| \;\frac{1}{|\mathrm{Im}\, z|^p}  \le c_p
  \  \hnorm{f}_m < \infty  \quad \text{for} \quad m \ge p+1
\end{equation}
with a constant $c_{p}$ (see \cite[Appendix B]{HuSi00} for details).

Thus we will pick appropriate smooth functions $f_{\pm}$ and estimate the
error between the quantity in \eqref{XPXP2} and the corresponding finite
volume quantity.  The error will be then controlled by the following lemma.

\begin{lemma}
  \label{inf-fin}
  Let ${I}\subset \Xi^{\mathrm{CL}} $ be a compact interval, so
  \eqref{aizenmolch} holds for all $E\in {I} $ with the same $\ell$ and $s$.
  Then there exists a constant $C $ such that for all $C^{4}$-functions
  $f_{\pm}$ with $\supp f_{\pm} \subset I$ and $\abs{f_{\pm}}\le1$, we have
  \begin{equation}
    \begin{split}\label{errorest}
      &   \big\lvert\EE\left\{ \langle\delta_{0}, F_{-} X_{1} F_{+}
        X_{1}F_{-} \delta_{0}\rangle  
        -  \langle\delta_{0}, F_{-,L} X_{1} F_{+,L}X_{1}F_{-,L}
        \delta_{0}\rangle \right\} \big\rvert \\ 
      & \hspace{1in}  \le C \bigl(1+ \hnorm{f_{-}}_3\bigr)^{\frac 2 3 }  \bigl(
        \hnorm{f_{+}}_4 +  \hnorm{f_{-}}_4\bigr)^{\frac 1 3} L^{\frac 4 3 d}
      \; \e^{-\frac {1} {12 \ell} L} 
    \end{split}  
  \end{equation}
  for all $L \ge 3$, where $F_{\pm}:= f_{\pm}(H)$ and $F_{\pm,L}:=
  f_{\pm}(H_{L})$.
\end{lemma}

\Proof  
  Since $ f _{\pm}= f_{\pm} \Chi_{I}$ and ${I}\subset \Xi^{\mathrm{CL}} $ with
  $I$ a compact interval, and $\abs{f_{\pm}}\le1$, it follows from
  \eqref{PdecayI} that
  \begin{equation}\label{Fdecay}
    \E \left \{ \left\lvert\langle\delta_{x},F_{\pm}
        \delta_{y}\rangle\right\rvert^{p}\right\} 
    \le C_ {I}\e^{-c_{I} \abs{x-y}} \quad  \quad \text{for all $p \in
      [1,\infty[$  and  $ x,y \in \ZZ^{d}$}, 
  \end{equation}
  where the constants $C_ {I}$ and $c_{I}>0$ are independent of $f _{\pm}$.
  The corresponding estimates for $F_{\pm,L}$ and $F_{\pm} -
  \widehat{F}_{\pm,L}$, the two main technical estimates needed for the proof
  of Lemma~\ref{inf-fin}, are isolated in the following sublemma.

  \begin{sublemma}
    \label{mainest} 
    Let the interval $I$ be as in Lemma~\ref{inf-fin}. Then there exist
    constants $C_{1}, C_{2} $ such that for all all $C^{4}$-functions $f$ with
    $\supp f \subset I$, $L \ge 3$, and all $x,y\in\ZZ^{d}$, we have
    \begin{align} \label{mainesti}
      \EE\bigl\{ |\langle\delta_{x}, (F -
      \widehat{F}_{L}) \delta_{y}\rangle | \bigr\}
      &  \le C_{1} \hnorm{f}_4 {L^{2d-2}}  \,\e^{- \frac {1} {2\ell } \{
        \dist(x,\partial 
        \Lambda_{L}) + \dist(y,\partial\Lambda_{L})\}}\\
      \intertext{and}
      \label{mainestii}
      \EE\bigl\{ |\langle\delta_{0}, 
      \widehat{F}_{L} \delta_{x}\rangle | \bigr\}
      & \le C_{2}  \hnorm{f}_3 {L^{d-1}} \;\e^{-  \frac {1} \ell
        |x|}\,\Chi_{\Lambda_{L}}(x), 
    \end{align}
    where 
    $F := f(H)$ and $\widehat{F}_{L} := f(\widehat{H}_{L})$. 
  \end{sublemma}

  \Proof
    Let $R(z):= (H - z)^{-1}$ and $\widehat{R}_{L}(z):= (\widehat{H}_{L} -
    z)^{-1}$ be the resolvents for $H$ and $\widehat{H}_{L}$. It follows from
    the resolvent identity that
    \begin{align}
      \widehat{R}_{L}(z)&=   R(z) + R(z)\Gamma_{L} \widehat{R}_{L}(z)\\
      &= \label{twiceresolv}
      R(z) + R(z)\Gamma_{L} R(z) -  R(z)\Gamma_{L}
      \widehat{R}_{L}(z)\Gamma_{L} R(z), 
    \end{align}
    where $\Gamma_{L }:=H- \widehat{H}_{L}$.  Note that either
    ${\Gamma_L(x,y)}=0$ or $\abs{\Gamma_L(x,y)}=1$, and if 
    $(x,y) \in \mathcal{E}_L := \{(x,y) \in \ZZ^d \times \ZZ^d : \Gamma_L(x,y) \neq
    0\} $ we must have either $x \in  \partial \Lambda_{L}$ or $y \in  \partial \Lambda_{L}$ (or both, we have periodic boundary condition), 
  and moreover $|\mathcal{E}_L| \le 8d^{2} L^{d-1}$.
   
    To prove \eqref{mainesti}, we first apply the Helffer--Sj\"ostrand formula
    \eqref{HS} to both $F$ and $\widehat{F}_{L}$, use \eqref{twiceresolv}, and
    the crude estimate $\| \widehat{R}_{L}(z)\| \le |\Im z|^{-1}$ to get
    \begin{equation} \label{istart}
    \begin{split}
      \EE \bigl\{ |& \langle \delta_{x}, (F -
      \widehat{F}_{L}) \delta_{y}\rangle | \bigr\} \\   
      & \le  \; |\mathcal{E}_L|  
      \sup_{(u,v) \in \mathcal{E}_{L}} \int_{\RR^2} \! 
      |\d\tilde{f}(z)| \; \EE\bigl\{ |R(z; x,u)|\,  |R(z; v,y)|  \bigr\}
      \\
      & \quad \;  +  |\mathcal{E}_L| ^{2} \sup_{\substack{(u,v) \in
          \mathcal{E}_{L}\\  (w',w) \in \mathcal{E}_{L}}} \int_{\RR^2} \! 
      |\d\tilde{f}(z)| \;|\Im z|^{-1}  \EE \bigl\{ |R(z; x,u)|\,    |R(z;
      w,y)|\bigr\}. 
    \end{split}
    \end{equation}
    We now exploit the crude bound $\| {R}(z)\| \le |\Im z|^{-1}$ and the
    Cauchy--Schwarz inequality to obtain fractional moments. This allows the
    use of \eqref{aizenmolch} for $\Re z \in \supp f \subset {I}\subset
    \Xi^{\mathrm{CL}} $, obtaining,
    \begin{equation}
      \label{useloc}
      \begin{split}
      \EE\bigl\{ |R(z; x,u)|&\, |R(z; v,y)|  \bigr\} \\ 
      & \le  |\Im z|^{s-2} \; \EE\{ |R(z; x,u)|^{s}\}^{\frac 1 2} 
      \EE\{|R(z; v,y)|^{s}\}^{\frac 1 2}  \\    
      & \le K |\Im z|^{s-2} \; \e^{- {\frac {1} {2\ell}}(|x-u| + |v-y|)}
    \end{split}
    \end{equation}
    for all $x,u,v,y \in \ZZ^{d}$.  Plugging the bound \eqref{useloc} into
    \eqref{istart}, and using \eqref{HShigherorder} and properties of the set
    $\mathcal{E}_L$, we get the estimate \eqref{mainesti}.
  
    The estimate \eqref{mainestii} is proved along the same lines.  We may
    assume $x\in\Lambda_{L}$, since otherwise the left hand side is clearly
    zero.  Proceeding as above, we get
    \begin{equation}
      \label{iichain}
      \begin{split}
        \EE\bigl\{ |\langle\delta_{0}, \widehat{F} \delta_{x}\rangle |\bigr\} 
            & \le \int_{\RR^2} \!
      |\d\tilde{f}(z)| \; \EE\{ |\widehat{R}_{L}(z; 0,x)| \} \\
      & \le \int_{\RR^2} \! 
      |\d\tilde{f}(z)| \; \EE\{ |{R}(z; 0,x)| \}  \\ 
      &  \qquad +
      |\mathcal{E}_L|\! \sup_{(u,v)\in\mathcal{E}_{L}} \int_{\RR^2} \!\! 
      |\d\tilde{f}(z)| \!\; |\Im z|^{-1}{\EE\{  |{R}(z; 0,u)| \}}
    \end{split}
    \end{equation}
    and 
    \begin{equation}
      \label{iiuseloc}
      \EE\{ |{R}(z; 0,x)| \} \le |\Im z|^{s-1} \;\EE\{  |{R}(z; 0,x)|^{s} \}
      \le   K  |\Im z|^{s-1} \,  \e^{- {\frac {1} \ell}|x|} .  
    \end{equation}
    The estimate \eqref{mainestii} now follows.
  \Endproof

  We may now finish the proof of  Lemma~\ref{inf-fin}.  We have
  \begin{equation}
    \langle\delta_{0}, F_{-,L} X_{1} F_{+,L}X_{1}
    F_{-,L}  \delta_{0}\rangle =\langle\delta_{0}, \widehat{F}_{-,L} X_{1}
    \widehat{F}_{+,L}X_{1} 
    \widehat{F}_{-,L}  \delta_{0}\rangle,
  \end{equation}
  since $\Chi_{\Lambda_{L}}F_{\pm,L} \Chi_{\Lambda_{L}}=
  \Chi_{\Lambda_{L}}\widehat{F}_{\pm,L} \Chi_{\Lambda_{L}}$ and the operators
  $F_{\pm,L}$ and $\widehat{F}_{\pm,L}$ commute with $ \Chi_{\Lambda_{L}}$.
  Thus
  \begin{align}\notag
    & \bigl|\EE\left\{ \langle\delta_{0}, F_{-} X_{1}  F_{+}
      X_{1}F_{-} \delta_{0}\rangle  
      -  \langle\delta_{0}, F_{-,L} X_{1} F_{+,L}X_{1}
      F_{-,L}  \delta_{0}\rangle \right\} \bigr|\\ \label{first}
    &  \qquad\qquad \le  \bigl|\EE\bigl\{ \langle\delta_{0},
    (F_{-}-\widehat{F}_{-,L}) X_{1} F_{+} 
    X_{1}F_{-} \delta_{0}\rangle  
    \bigr\} \bigr|\\\label{second}
    & \qquad\qquad\quad +\bigl|\EE\bigl\{ \langle\delta_{0},
    \widehat{F}_{-,L} X_{1} F_{+} 
    X_{1}(F_{-}-\widehat{F}_{-,L})\delta_{0}\rangle  
    \bigr\} \bigr|\\\label{third}
    & \qquad\qquad\quad + \bigl|\EE\bigl\{ \langle\delta_{0},
    \widehat{F}_{-,L} X_{1} (F_{+}-\widehat{F}_{+,L})    X_{1}
    \widehat{F}_{-,L} \delta_{0}\rangle   
    \bigr\} \bigr|.
  \end{align}  
  Each term in the above inequality  can be estimated by H\"older's inequality:
  \begin{equation}
    \label{AAA} 
    \begin{split}
    &\left|\EE\left\{ \langle\delta_{0}, A_{1} X_{1} A_{2}
        X_{1}A_{3} \delta_{0}\rangle  
      \right\} \right|\\ 
    & \quad
    \le \sum_{x,y\in\ZZ^{d}} |x_{1}| \,|y_{1}| \,
    \EE\left\{ |A_{1} (0,x)| \,
      |A_{2}(x,y)| \, |A_{3}(y,0)| \right\} \\ 
    & \quad
    \le \sum_{x,y \in\ZZ^{d}} |x_{1}| \,|y_{1}| \,
    \EE\left\{ |A_{1} (0,x)|^{3}\right\}^{\frac 1 3}  \EE\left\{
      |A_{2}(x,y)|^{3}\right\}^{\frac 1 3}    \EE\left\{
      |A_{3}(y,0)|^{3}\right\}^{\frac 1 3} ,
  \end{split}
  \end{equation}
  where $A_{j}$, $j=1,2,3$, may be either $F_{\pm}$, $\widehat{F}_{-,L}$, or
  $F_{\pm}-\widehat{F}_{\pm,L}$.  We estimate $ \EE\left\{|
    F_{\pm}(x,y)|^{3}\right\}$ by \eqref{Fdecay} and $ \EE\bigl\{|
  \widehat{F}_{-,L}(0,x)|^{3}\bigr\}$ by \eqref{mainestii}. If follows from
  \eqref{mainesti} that
  \begin{align}\notag
    &\EE\bigl\{|(F_{-}-\widehat{F}_{-,L}) (0,x)|^{3}\bigr\} \le  4 \, \EE
    \bigl\{ \abs{(F_{-}-\widehat{F}_{-,L}) (0,x)}\bigr\}\\\label{FFxy} 
    &\qquad  \qquad \le 4 C_{1}   \hnorm{f_{-}}_4 {L^{2d-2}} \,\e^{- \frac
      {1}{2 \ell} ( \dist(0,\partial 
      \Lambda_{L}) + \dist(x,\partial\Lambda_{L}))}\\
    & \qquad \qquad  \le 4 C_{1}   \hnorm{f_{-}}_4 {L^{2d-2}}\,\e^{- \frac
      {1}{2 \ell}  \frac {L-3} 2}, 
  \end{align}
  since $ \abs{(F_{-}-\widehat{F}_{-,L}) (0,x)}\le 2$ and $\dist(0,
  \partial\Lambda_L) \ge \frac {L-3} 2$. Thus we get, with some constant $C$,
  \begin{equation}  \label{firstsecond}
    {\eqref{first}} + {\eqref{second}} 
    \le  C \bigl(1+ \hnorm{f_{-}}_3^{\frac 1 3}\bigr) \hnorm{f_{-}}_4^{\frac 1
      3} {L^{d-1}}\,\e^{- \frac {1} {6\ell} \frac {L-3} 2}. 
  \end{equation}
  To estimate \eqref{third}, we control $\EE\bigl\{|(F_{+}-\widehat{F}_{+,L})
  (x,y)|^{3} \bigr\}$ from \eqref{mainesti} as in \eqref{FFxy}. We get, with
  constant $C^{\prime}$,
  \begin{equation}
    \label{thirdest} 
    \begin{split}
    \eqref{third} & \le
    C^{\prime}  {L^{\frac 4 3(d-1)}} \hnorm{f_{-}}_3^{\frac 2 3} 
    \hnorm{f_{+}}_4^{\frac 1 3} \\
    & \qquad \times \sum_{x,y \in\Lambda_{L}} |x_{1}| \,|y_{1}|  \,
    \e^{- \frac {1} {3\ell} ( \abs{x} + \abs{y})} 
    \,\e^{- \frac {1} {6\ell} ( \dist(x,\partial
      \Lambda_{L}) + \dist(y,\partial\Lambda_{L}))}\\
    &  \le C^{\prime}  {L^{\frac 4 3(d-1)}} \hnorm{f_{-}}_3^{\frac 2 3} 
    \hnorm{f_{+}}_4^{\frac 1 3}  \,\e^{- \frac {1} {6\ell} \frac {L-3}
      2},
  \end{split}
  \end{equation}
  since for $x \in \Lambda_{L}$ we have
  \begin{equation}
    \abs{x} + \dist(x,\partial  \Lambda_{L})\ge \dist(0,\partial
    \Lambda_{L})\ge  \tfrac {L-3}2.
  \end{equation}
  
  The desired estimate \eqref{errorest} now follows from
  \eqref{first}--\eqref{third}, \eqref{firstsecond}, and \eqref{thirdest},
  with a suitable constant $C$.  \Endproof

%
\Subsec{The finite volume estimate}
\label{thirdpart}
For the finite volume Anderson Hamiltonian $H_{L}$ we have available a
beautiful estimate due to Minami \cite{Min96}, which may be stated as
\begin{equation}\label{minami2}
  \E \bigl\{ \{ \tr \Chi_{I}(H_{L}) \}^{2} -  \tr \Chi_{I}(H_{L})
  \bigr\}\le  
  \pi^2  \|\rho\|_\infty^{2} |I|^2 |\Lambda_{L}|^{2}
\end{equation}
for all intervals $I\subset \RR$ and length scales $L \ge 1$.  (See
\cite[Appendix~A]{KlMo05} for an outline of the argument.)  Although Minami
wrote his original proof for Dirichlet boundary condition, the result is valid
for the usual boundary conditions, and in particular for periodic boundary
condition.

\begin{remark}
  \label{missingpower}  
  The dependence on $L \sim |\Lambda_{L}|^{\frac 1 d}$ in the right hand side
  of \eqref{minami2} is optimal; it cannot be improved.  Ergodicity implies
  that
  \begin{equation}
    \lim_{L \to \infty} \tfrac 1 {|\Lambda_{L}|}  \tr \Chi_{B}(H_{L}) = \EE
    \left\{ \langle    \delta_{0}, \Chi_{B}(H)  \delta_{0} \rangle  \right\}=
    \mathcal{N}(B) \quad \text{$\PP$-a.s.}, 
  \end{equation}
  where $ \mathcal{N}(B)$ is the density of states measure.  If $I$ and
  $I_{\pm}$ are intervals of nonzero lengths contained in the spectrum of $H$,
  we must have $ \mathcal{N}(I),\mathcal{N}(I_{\pm})>0$, and hence
  \begin{align}\label{minami25}
    \lim_{L \to \infty} \tfrac 1 {|\Lambda_{L}|^{2}} \;\E \bigl\{ \{ \tr
    \Chi_{I}(H_{L}) \}^{2} -  \tr \Chi_{I}(H_{L}) \bigr\} &=
    \mathcal{N}(I)^{2 }>0,\\ 
    \lim_{L \to \infty} \tfrac 1 {|\Lambda_{L}|^{2}} \;\E \bigl\{ \{ \tr
    \Chi_{I_{+}}(H_{L}) \} \{ \tr \Chi_{I_{-}}(H_{L}) \}
    \bigr\} &=  \mathcal{N}(I_{+}) \mathcal{N}(I_{-})>0. 
  \end{align}
\end{remark}

\begin{lemma}
  \label{minami}
  Let $J_{\pm}\subset \RR$ be intervals such that $J_{-} \cap
  J_{+}=\emptyset$, and consider an interval $J \supset J_{-} \cup J_{+}$.
  Then, given Borel functions $f_{\pm}$ on $\RR$ with $0\le
  {f_{\pm}}\le\Chi_{J_{\pm}}$, we have
  \begin{equation}\label{finitevolumeMott}
    \EE \bigl\{ \langle\delta_{0}, F_{-,L} X_{1} F_{+,L}X_{1}F_{-,L}
    \delta_{0}\rangle  \bigr\} \le   \tfrac{ \pi^2} 4   \|\rho\|_{\infty}^{2}
    |J|^{2} L^{d+2} 
  \end{equation}
  for all $L\ge 3$, where $F_{\pm,L}= f_{\pm}(H_{L})$.
\end{lemma}

\Proof 
  With periodic boundary condition, finite volume expectations are invariant
  with respect to translations (in the torus). This, combined with $F_{-,L}
  F_{+,L}=0$, gives
  \begin{equation}
    \label{minstart}
    \begin{split}
    \EE \bigl\{ \langle\delta_{0}, & F_{-,L} X_{1}  F_{+,L}X_{1}F_{-,L}
    \delta_{0}\rangle  \bigr\}  \\
     & =  
    \tfrac 1 { |\Lambda_L|} \sum_{x\in\Lambda_{L}} \EE \bigl\{
    \langle\delta_{x}, F_{-,L} X_{1} F_{+,L}X_{1}F_{-,L} \delta_{x}\rangle
    \bigr\}  \\
     & =  \tfrac 1 { |\Lambda_L|} \; \EE \bigl\{ \tr \{ F_{-,L} X_{1}
    F_{+,L}X_{1}F_{-,L} \}\bigr\} ,   
  \end{split} 
  \end{equation}
  where the trace is taken in $\ell^{2}(\Lambda_{L})$.  Since $\norm{
    X_{1,L}}\le \frac L 2$, where $X_{1,L}=X_{1}\Chi_{\Lambda_{L}}$ is the
  restriction of $X_{1}$ to $\ell^{2}(\Lambda_{L})$, $0 \le F_{\pm,L}\le
  Q_{\pm,L}:= \Chi_{J_{\pm}}(H_{L})$, and $Q_{+,L} +Q_{-,L}\le Q_{L}:=
  \Chi_{J}(H_{L}) $, we have
  \begin{align} \label{FtoQ1}
    &\tr \bigl\{ F_{-,L} X_{1} F_{+,L}X_{1}F_{-,L} \bigr\} \le 
    \norm{X_{1,L} F_{+,L}X_{1,L}}
    \bigl(\tr  F_{-,L} ^{2} \bigr)\\
    & \qquad \qquad \label{FtoQ}
    \le  \tfrac {L ^{2}} 4  \norm{ F_{+,L}}
    \bigl(\tr  F_{-,L} ^{2} \bigr)  \le  \tfrac {L ^{2}} 4\bigl(\tr  F_{+,L}
    \bigr)\bigl(\tr  F_{-,L} ^{2} \bigr) \\ 
    & \qquad \qquad \le  \tfrac {L ^{2}} 4   \bigl(\tr  Q_{+,L}  \bigr)^{}
    \bigl(\tr  Q_{-,L}  \bigr) \le \tfrac {L ^{2}} 4 \bigl\{   \bigl(\tr  Q_{L}
    \bigr)^{2}- \tr  Q_{L}   
    \bigr\}. \label{FtoQ2}
  \end{align}
  Combining \eqref{minstart} and \eqref{FtoQ1} -- \eqref{FtoQ2}, and using
  Minami's estimate \eqref{minami2}, we get
  \begin{equation}\label{powerofL}
    \EE \bigl\{ \langle\delta_{0}, F_{-,L} X_{1} F_{+,L}X_{1}F_{-,L}
    \delta_{0}\rangle  \bigr\}  \le   \tfrac{ \pi^2} 4   \|\rho\|_\infty^{2}
    |J|^2 |\Lambda_{L}| {L ^{2}} , 
  \end{equation}
  which yields \eqref{finitevolumeMott}.
\Endproof

%
\Subsec{The  proof of Theorem~\ref{ourmott2}}  
\label{combine}
We now have all the ingredients to prove Theorem~\ref{ourmott2}.  Since $E_{F}
\in \Xi^{\mathrm{CL}}$, there is $\nu_{0}\in]0,1[$ such that
$I_{0}:=[E_{F}-\nu_{0}, E_{F}+\nu_{0}] \subset \Xi^{\mathrm{CL}}$, and
\eqref{aizenmolch} holds for all $E\in {I}_{0} $ with the same exponent
$s=s_{E_{F}}$ and localization length $\ell=\ell_{E_{F}}$. Given
$\nu\in]0,\nu_{0}]$, we define compact intervals
\begin{equation}
  \begin{split}
    I&:=[E_{F}-\nu, E_{F}+\nu],\\
    I_{-}& := ]E_{F} - \nu, E_{F}] \quad\text{and}\quad  
    J_{-} := ]E_{F} - \nu +{\nu^{4}}, E_{F} -  {\nu^{4}}]  , \\
    I_{+} &:= ]E_{F}, E_{F} + \nu]    \quad\text{and}\quad  J_{+} := ]E_{F} +
    \nu^{4}, E_{F} + \nu - \nu^{4}] . 
  \end{split}
\end{equation}
Note that $J_{\pm} \subset I_{\pm} \subset I \subset \Xi^{\mathrm{CL}}$ and
$I_{-} \cap I_{+} = \emptyset$.  Moreover, we have $ I_{\pm }\backslash
J_{\pm}= J_{\pm,1}\cup J_{\pm,2}$, where $J_{\pm,j}$, $j=1,2$, are intervals
of length $|J_{\pm,j}|=\nu^{4} $.  Thus
\begin{align}\notag
  \hspace*{1.2cm}
  \Psi_{E_{F}}(I_{+}\times I_{-})  & =  \Psi_{E_{F}}(J_{+}\times J_{-})
  +  \Psi_{E_{F}}(I_{+}\times J_{-,1}) + \Psi_{E_{F}}(I_{+}\times
  J_{-,2})\\   
  & \qquad  \qquad \label{4error2} 
    +  \Psi_{E_{F}}(J_{+,1}\times J_{-})+ \Psi_{E_{F}}(J_{+,2}\times
  J_{-})\\ 
  &\le   \Psi_{E_{F}}(J_{+}\times J_{-}) + 4W_{\frac 1 2}\;
  \nu^{2}, \notag
\end{align}
where the four terms containing $J_{\pm,1}$ and $J_{\pm,2}$ were estimated by
Lemma~\ref{wegner} (with $\beta =1/2$).
  
To estimate $ \Psi_{E_{F}}(J_{+}\times J_{-})$, we exploit the existence
of $C^{4}$-functions $f_{\pm}$ such that $\Chi_{J_{\pm}} \le f_{\pm} \le
\Chi_{I_{\pm}}$ and $|f_{\pm}^{(k)}| \le 2 \nu^{-4k}\Chi_{I_{\pm}\backslash
  J_{\pm}} $, $k=1,2,3,4$.  Note that
\begin{equation}\label{fnu}
  \hnorm{f_{\pm}}_3 \le \hnorm{f_{\pm}}_4\le C \nu^{-16}\nu= C\nu^{-15},
\end{equation}
where the constant $C$ is independent of $\nu \in ]0,\nu_{0}]$ and $f_{\pm}$.
Using first \eqref{XPXP2} in Lemma~\ref{dictionary2} (with $B_{\pm} = J_{\pm}$
and $F_{\pm}=f_{\pm}(H)$) to replace the spectral projections by smooth
functions of $H$, followed by Lemma~\ref{inf-fin} to achieve the passage to
finite volume, we get
\begin{equation}
  \begin{split}\label{tofinitevolume}
    ~\Psi_{E_{F}}(J_{+}\times J_{-}) & \le   \EE \bigl\{ \langle\delta_{0},
    F_{-} X_{1}F_{+} X_{1} F_{-} \delta_{0} \rangle\bigr\}\\ 
    &  \le  \EE\bigl\{ \langle\delta_{0}, F_{-,L} X_{1}
    F_{+,L}X_{1}F_{-,L} \delta_{0}\rangle \bigr\} + C^{\prime} \nu^{{-15}}
    L^{\frac 4 3 d} \; \e^{-\frac {1} {12 \ell} L}
  \end{split}
\end{equation}
for all $L \ge 3$,  where $F_{\pm,L}= f_{\pm}(H_{L})$.

Combining \eqref{4error2} and \eqref{tofinitevolume}, and using
Lemma~\ref{minami} to estimate the finite volume quantity, we get
\begin{align}
  \Psi_{E_{F}}(I_{+}\times I_{-})  \le 
  { \pi^2}   \|\rho\|_{\infty}^{2} \nu^{2} L^{d+2}+ 
  C^{\prime} \nu^{{-15}}  L^{\frac 4 3 d} \;
  \e^{-\frac {1} {12\ell} L} +4W_{\frac 1 2} \nu^{2}.
\end{align}
If we now choose
\begin{equation}
  \label{Lchoice}
  L = (17\cdot12  +1) { \ell}  \,\log \tfrac{1}{\nu}= 205{ \ell}  \,\log
  \tfrac{1}{\nu} , 
\end{equation}
then there exists $\nu_{0}^{\prime}\in ]0,\nu_{0}]$, such that for all $\nu
\in ]0,\nu_{0}^{\prime}]$ we have
\begin{align}\label{finalest}
  \Psi_{E_{F}}(I_{+}\times I_{-})  \le 205^{d+2}
  { \pi^2}   \|\rho\|_{\infty}^{2} \ell^{d +2} \nu^{2} \left(\log
    \tfrac{1}{\nu}\right)^{d+2}+  
  C^{\prime\prime}  \nu^{2},
\end{align}
from which \eqref{leadingboundN} follows.

Theorem~\ref{ourmott2} is proven, yielding Theorem~\ref{ourmott}.

\begin{remark} \label{missingpower2} 
  As discussed in Remark~\ref{remMot}, in our estimate for Mott's formula,
  namely \eqref{leadingbound8} (or, equivalently,  \eqref{leadingboundN}), the exponent of
  $\log\frac 1 \nu$ is $d+2$, instead of $d+1$ as in \eqref{mott}.  This comes
  from \eqref{powerofL}, where we get a factor of $L^{d+2}$.  As seen in
  Remark~\ref{missingpower}, the power of $L$ we acquire in the passage from
  \eqref{FtoQ2} to \eqref{powerofL} cannot be improved. The factor of $L^{2}$
  obtained going from \eqref{FtoQ1} to \eqref{FtoQ} must also be correct
  because of \eqref{Lchoice}, since we need $L^{d+2}$ in \eqref{powerofL} to
  get $\ell^{d+2}$ in \eqref{finalest}.  To obtain a factor of $(\log\frac 1
  \nu)^{d+1}$ as in \eqref{mott}, we would need to improve the estimate in
  \eqref{FtoQ1}--\eqref{FtoQ} to gain an extra factor of $(\log\frac 1
  \nu)^{-1}$.  This seems far-fetched to us.
\end{remark}

\begin{remark}\label{lowerrem}  
  Starting from the lower bound given in Proposition~\ref{sigmaUpsilon}, and
  proceeding as in the derivation of \eqref{tofinitevolume}, we obtain the
  lower bound
  \begin{equation}\label{lowerbsigma}
    \overline{\sigma}_{E_{F}}^{\mathrm{in}}(\nu) \ge \tfrac \pi  2  \EE\bigl\{
    \langle\delta_{0}, G_{-,L} X_{1} G_{+,L}X_{1}G_{-,L} \delta_{0}\rangle
    \bigr\} + C^{\prime\prime} \nu^{{-15}}  L^{4d} \; 
    \e^{-\frac {1} {12 \ell} L},
  \end{equation}
  where $G_{\pm,L} := g_{\pm}(H_{L})$ and the functions $g_{\pm}$
  satisfy $\Chi_{B_{\pm}} \le g_{\pm} \le \Chi_{J_{\pm}}$ with
  $J_{\pm}$ as in \eqref{I+I-}, $B_- := ]E_{F} - \tfrac \nu 2 +
  \nu^{4}, E_{F} -\tfrac \nu 4 - \nu^{4}]$, and $ B_+ := ]E_{F} +
  \tfrac \nu 4 + \nu^{4}, E_{F} + \tfrac \nu 2 - \nu^{4}]$. Moreover,
  the functions $g_{\pm}$ are supposed to satisfy the hypotheses of
  Lemma~\ref{inf-fin} with respect to the intervals $B_{\pm}$, the
  estimate \eqref{fnu}, and $f_{\pm}=\sqrt{g_{\pm}}$ satisfy the
  hypotheses of Lemma~~\ref{dictionary2} with respect to the intervals
  $B_{\pm}$.  Unfortunately, we are not able to obtain a useful lower
  bound for $\overline{\sigma}_{E_{F}}^{\mathrm{in}}(\nu)$ from
  \eqref{lowerbsigma} because we do not have a lower bound for the
  finite volume term; Minami's estimate gives only an upper bound.
\end{remark}

%


\end{document}